\documentclass[pra,twocolumn,showpacs,amsmath,amssymb]{revtex4-1}

\usepackage{graphicx}
\usepackage{amsfonts}
\usepackage{amsmath}
\usepackage{amssymb}
\usepackage{color}
\usepackage{bm}
\usepackage{bbm}
\usepackage{enumerate}
\usepackage{color}
\graphicspath{figures} 
\usepackage{amsfonts, amsmath, amsthm, amssymb} 
\usepackage{mathtools}
\usepackage{array}
\usepackage[colorlinks,bookmarks=false,citecolor=blue,linkcolor=red,urlcolor=blue]{hyperref}

\usepackage[colorlinks,bookmarks=false,citecolor=blue,linkcolor=red,urlcolor=blue]{hyperref}
\usepackage[usenames,dvipsnames,svgnames,table]{xcolor}
\graphicspath{figures} 

\newcommand{\be}{\begin{equation}}
\newcommand{\bee}{\begin{equation*}}
\newcommand{\ee}{\end{equation}}
\newcommand{\eee}{\end{equation*}}
\newcommand{\bearre}{\begin{eqnarray*}}
\newcommand{\eearre}{\end{eqnarray*}}
\newcommand{\bearr}{\begin{eqnarray}}
\newcommand{\eearr}{\end{eqnarray}}

\begin{document}

\title{
	Entanglement and fluctuations in the XXZ model with power-law interactions
}

\author{Ir\'en\'ee Fr\'erot$^1$
\footnote{Electronic address: \texttt{irenee.frerot@ens-lyon.fr}},
	Piero Naldesi $^{2,3}$
 and Tommaso Roscilde$^{1,4}$
 \footnote{Electronic address: \texttt{tommaso.roscilde@ens-lyon.fr}}
 }

\affiliation{$^1$ Univ Lyon, Ens de Lyon, Univ Claude Bernard, CNRS, Laboratoire de Physique, F-69342 Lyon, France}
\affiliation{$^2$ Dipartimento di Fisica e Astronomia dell'Universit\`a di Bologna, Via Irnerio 46, 40127 Bologna, Italy}
\affiliation{$^3$ INFN, Sezione di Bologna, Via Irnerio 46, 40127 Bologna, Italy}
\affiliation{$^4$ Institut Universitaire de France, 103 boulevard Saint-Michel, 75005 Paris, France}
\date{\today}


\begin{abstract}
{We investigate the ground-state properties of the spin-$1/2$ XXZ model with power-law-decaying ($1/r^{\alpha}$) interactions, describing spins interacting with long-range transverse (XX) ferromagnetic interactions and longitudinal (Z) antiferromagnetic interactions, or hardcore bosons with long-range repulsion and hopping. The long-range nature of the couplings allows us to quantitatively study the spectral, correlation and entanglement properties of the system by making use of linear spin-wave theory, supplemented with density-matrix renormalization group in one-dimensional systems. Our most important prediction is the existence of \emph{three} distinct coupling regimes, depending on the decay exponent $\alpha$ and number of dimensions $d$:  
1) a short-range regime for $\alpha > d + \sigma_c$ (where $\sigma_c = 1$ in the gapped N\'eel antiferromagnetic phase exhibited by the XXZ model, and $\sigma_c = 2$ in the gapless XY ferromagnetic phase), sharing the same properties as those of finite-range interactions ($\alpha=\infty$); 2) a long-range regime $\alpha < d$, sharing the same properties as those of the infinite-range interactions ($\alpha=0$) in the thermodynamic limit; and 3) a most intriguing medium-range regime for $d < \alpha < d+\sigma_c$, continuously interpolating between the finite-range and the infinite-range behavior. The latter regime is characterized by elementary excitations with a long-wavelength dispersion relation $\omega \approx \Delta_g + ck^z$ in the gapped phase, and $\omega \sim k^z$ in the gapless phase, exhibiting a continuously varying dynamical exponent $z = (\alpha - d) / \sigma_c$. In the gapless phase of the model the $z$ exponent is found to control the scaling of fluctuations, the decay of correlations, and a universal sub-dominant term in the entanglement entropy, leading to a very rich palette of behaviors for ground-state quantum correlations beyond what is known for finite-range interactions.}
\end{abstract}

\maketitle

\section{Introduction}

Long-range (LR) interacting quantum many-body systems have attracted an increasing level of attention 
in the recent years. The experimental improvements in cooling, controlling and addressing few- to many-body atomic and molecular quantum systems possessing sizable LR interactions \cite{Lahayeetal2009,Schneideretal2012,Monroeetal2014,Browaeysetal2016} have triggered intense theoretical efforts, aimed at exploring the possibility that such interactions may stabilize stronger quantum collective phenomena with respect to the case of short-range interactions. In trapped-ion experiments, in particular, it has become possible to engineer Ising and exchange interactions between the spin of the ions decaying as a power-law $1/r^\alpha$ of the distance $r$, and with a continuously tunable exponent $\alpha$ ($0<\alpha <3$ \cite{brittonetal2012}).
Within this setup, experimentalists were able to observe how the dynamics of correlation spreading after a quantum quench is modified by the long-range interactions with respect to the case of ultracold neutral atoms interacting via a contact potential \cite{cheneauetal2012}. 
Within the context of ultracold neutral gases several groups have attained the quantum degeneracy of atoms possessing a large intrinsic magnetic moment,  namely fermionic and bosonic isotopes of Cr \cite{griesmaieretal2005, dePazetal2013, nayloretal2015}, Dy \cite{luetal2011, luetal2012} and Er \cite{aikawaetal2012, aikawaetal2014}, and they were able to observe the coherent spin-exchange dynamics in these systems, induced by the large dipole-dipole ($1/r^3$) interaction \cite{dePazetal2013, baieretal2016}. Moreover infinite-range cavity-mediated interactions in a Bose-Einstein condensate have been experimentally demonstrated \cite{Baumannetal2010,Ritschetal2013}, leading to the spontaneous formation of long-range ordered phases (solid and supersolid).
Finally, recent experimental progress in the manipulation of molecular systems with a large electric dipole \cite{hazzardetal2013, boetal2013} and of ensembles of Rydberg atoms \cite{schaussetal2015, barredoetal2015,Browaeysetal2016} has opened new pespectives for the quantum simulation of spin Hamiltonians with LR couplings.

In parallel to these remarkable experimental achievements, the theoretical efforts have focused on the study of equilibrium and out-of-equilibrium properties of LR interacting quantum lattice models.
A large number of works has focused on the peculiar post-quench spreading of correlations \cite{eisertetal2013, haukeT2013, hazzardetal2014, schachenmayeretal2015, foss-feigetal2015, cevolanietal2015, maghrebietal2016, buyskikhetal2016, cevolanietal2016} and entanglement \cite{schachenmayeretal2013, haukeT2013, hazzardetal2014, buyskikhetal2016} in LR interacting systems, in connection with the breakdown (or generalizations) of Lieb-Robinson bounds \cite{liebR1972} constraining the dynamics \cite{eisertetal2013, foss-feigetal2015}. Further studies have rather focused on the ground-state properties of these models, where LR interactions can strongly affect the decay of correlation functions \cite{duttaB2001, laflorencieetal2005, peteretal2012, vodolaetal2014}, lead to phase transitions \cite{duttaB2001, capogrosso-sansoneetal2010, maghrebietal2015, gongetal2016} or modify substantially the entanglement properties \cite{vodolaetal2014}. In particular, the breakdown of the Mermin-Wagner theorem for sufficiently LR interactions leads to the possibility of spontaneous breaking of a continuous symmetry even in one-dimensional systems \cite{duttaB2001, maghrebietal2015, gongetal2016}.

 In the face of the mounting body of experimental and theoretical results on LR interacting systems, it is of central importance to develop a broad (and possibly exhaustive) picture of the effect of LR interactions on the many-body physics of the system, and particularly so as the strength and decay law of interactions is varied continuously. While this endeavor might be too arduous to pursue for all the models of interest to experiment and theory in recent years, one can adopt a different strategy which focuses on a sufficiently simple model possessing nonetheless a rich phenomenology, and aims at extracting the most salient features and driving principles - in terms of excitations, fluctuations and entanglement - of the various regimes of LR interactions. 
 
  \begin{table*}[ht!]
	\begin{tabular}{| m{4cm} |  c  | c | c |}
	\toprule
	  {\bf XY phase} &  ~SHORT RANGE ~($\alpha > d+2$) &  ~ MEDIUM RANGE~ ($d < \alpha < d+2$) & ~LONG RANGE ~($\alpha < d$) \\
	 \hline
	 dynamical exponent $z$ ~~~ ($\omega \sim k^z$)  \null & $z=1$ &  $z=(\alpha - d) / 2$ & $z=0$ \\
	 \hline 
	   $\langle (\delta S^x)^2\rangle$ scaling (total ~~~~system) \null & $L^{\max(d, 2)}$ & $L^{\max(d,2z)}$ &  $L^d$   \\ 
	 \hline 
	   $\langle (\delta S^y)^2\rangle$ scaling (total~~~~ system) \null & $L^{d+1}$ & $L^{d+z}$ &  $L^d$ \\ 
	 \hline 
	   $\langle (\delta S^z)^2\rangle$ scaling (on the $A$ subsystem) \null & $L_A^{d-1} \log L_A$ & $L_A^{d-z}$ &  $L_A^d$  \\ 
	 \hline
	  scaling of subsystem entanglement entropy: ~~~ dominant term & $L_A^{d-1} $ ($\log L_A$ in $1d$)  & $L_A^{d-1}$ ($\log L_A$ in $1d$) &  ---  \\ 
	 \hline
	  scaling of subsystem entanglement entropy: ~~~~{ logarithmic term} & $\frac{d-1}{2}~ \log L_A$ & $ \frac{d-z}{2} ~\log L_A$ &  $\frac{d}{2} \log L_A$ \\ 
	 \hline
	 \end{tabular}
	 
	 \bigskip
	 
	 \begin{tabular}{| m{4cm} |  c  | c | c |}
	\toprule
	  {\bf N\'eel phase} &  ~SHORT RANGE ~($\alpha > d+1$) &  ~ MEDIUM RANGE~ ($d < \alpha < d+1$) & ~LONG RANGE ~($\alpha < d$) \\
	 \hline
	 dynamical exponent $z$ ($\omega \sim \Delta_g + c k^{z}$)  \null & { $z=\min(2, \alpha-d)$} &  $z=\alpha - d$ & --- \\
	 \hline 
	   $\langle (\delta S^{x(y)})^2\rangle$ scaling (total ~~~~ system) \null & $L^{d}$ & $L^{d}$ &  --- \\ 
	 \hline 
	   $\langle (\delta S^z)^2\rangle$ scaling (on a the $A$ subsystem) \null & $L_A^{d-1}$ & $L_A^{d-1}$ &  --- \\ 
	 \hline
	  scaling of subsystem entanglement entropy: ~~~ dominant term & $L_A^{d-1} $ & $L_A^{d-1}$ &  --- \\ 
	 \hline
	 \end{tabular}

	\caption{ \label{table_summary} Summary of main results for the long-range ordered XY phase (in $d=2$ and 3 for all $\alpha$, and in $d=1$ for $\alpha < 3$) and the N\'eel phase, concerning the dispersion relation, the scaling of fluctuations and of entanglement. The scaling laws refer to a hypercubic-lattice geometry with linear size $L$, or to a $A$ subsystem with linear extent $L_A$. The `---' symbols indicate the absence of predictions in the corresponding regimes: in the XY phase for $d=1$ the area-law term in the scaling of the entanglement entropy becomes a subdominant constant; and the N\'eel phase is simply absent for $\alpha<d$ (see the phase diagram on Fig. \ref{f.phase_diagram_2d}).
	}
	\end{table*}

 In this work we pursue the latter strategy by focusing our attention on the ground-state physics of a paradigmatic model in the theory of magnetism, the XXZ model for $S=1/2$, possessing power-law decaying ($1/r^{\alpha}$) isotropic ferromagnetic interactions for the $x$ and $y$ spin components, and ferro- or antiferromagnetic interactions for the $z$ spin components. The LR XXZ model describes also the physical situation of hardcore bosons with long-range hopping and density-density interactions. This model offers a rich showcase of effects of LR interactions on \emph{symmetry breaking phases}: 1) in any physical dimension $d$ its phase diagram exhibits an extended \emph{ferromagnetic XY phase} with breaking of the continuous rotation symmetry in the $xy$ plane, and the apparition of a gapless Goldstone mode. In particular in $d=1$ such a phase exists only thanks to the LR interactions, and it is stabilized by a decay exponent $\alpha<3$ against a Luttinger-liquid (namely gapless disordered) phase; 2) the above phase is competition with a \emph{N\'eel phase} exhibiting long-range antiferromagnetic ordering  along the $z$ axis and a gapped spectrum. As a consequence the LR XXZ model allows to monitor the effect of LR interactions with a continuously varying exponent $\alpha$ on phases breaking either a continuous or a discrete symmetry. In the XY phase the LR interactions are expected to stabilize against fluctuations the long-range order which already appears for nearest-neighbor interactions (in $d\geq 2$) . This fully justifies to treat such fluctuations as harmonic ones, as done by linear spin-wave (LSW) theory, on which we base most of our analysis. On the other hand the N\'eel phase is destabilized by the LR interactions due to their frustrated nature. Nonetheless in both phases, as well as at the transition between the two, the assumption of weak quantum fluctuations underlying LSW theory is well verified a posteriori, making our predictions quantitative. The case of antiferromagnetic XY interactions could also be treated in principle by LSW theory -- but in this case the power-law decay induces frustration, which is expected to progressively weaken long-range order, making the LSW approach less and less justified as $\alpha$ decreases.
 
  In particular LSW theory puts on the center stage the dispersion relation of free bosonic excitations, which is found in turn to control the scaling of fluctuations (or, equivalently, the decay of correlations), and the scaling of entanglement.  In particular a systematic analysis of the long-wavelength properties of the dispersion relation allows to identify \emph{three distinct regimes} upon varying the $\alpha$ exponent: 1) a \emph{short-range} regime for $\alpha > d+\sigma_c$ ($\sigma_c=1$ in the N\'eel phase and $\sigma_c=2$ in the XY phase), whose main properties reproduce those observed in the limit $\alpha =\infty$ of finite-range interactions; 2) a \emph{medium-range} regime for $d < \alpha < d+\sigma_c$, whose scaling properties (in the XY phase) are dominated by a continuously varying dynamical exponent $z = (\alpha-d)/\sigma_c$, governing the $k\to 0$ limit of the dispersion relation; 3) and a properly defined \emph{long-range} regime for $\alpha < d$, possessing dispersionless excitations, and reproducing the properties of the infinite-range limit $\alpha = 0$ in the thermodynamic limit. In particular this classification shows that the most interesting regime is the intermediate (medium-range) one, as it is the only one with markedly distinct features with respect to the two extreme limits of finite-range ($\alpha = \infty$) and infinite-range ($\alpha=0$) interactions which have been thoroughly investigated in the past. Importantly, the medium-range regime of the XY phase is also the one in which to frame the dipolar interaction in $d=$1, 2 and 3 (although $\alpha = 3$ falls on the boundaries of the medium-range regime for $d=1$ and 3, and in its bulk only for $d=2$). 
  
      Despite their harmonic nature within LSW theory, ground-state quantum fluctuations exhibit a very rich structure in terms of scaling properties, and they are associated with an equally complex scaling of the entanglement entropy (EE) of a subsystem. A list of such properties in the three regimes of the LR XXZ model is presented in Table~\ref{table_summary}, which summarizes the most important results of the present work.     
      
      The structure of our paper is as follows: Sec.~\ref{s_Ham} introduces the model and is theoretical treatment; Secs.~\ref{s_quantum_correlations} and ~\ref{s_entanglement} discuss the structure of quantum fluctuations and entanglement in the ground state, respectively; conclusions are drawn in Sec.~\ref{s_conclusions}.

\section{The Hamiltonian and its low energy properties.}
\label{s_Ham}
	In this Section, we introduce the model Hamiltonian under investigation, and proceed to determine its phase diagram within mean-field theory. We then investigate harmonic quantum fluctuations around the mean-field limit by diagonalizing the quadratic (LSW) Hamiltonian, and obtaining the dispersion relation of elementary excitations. The behavior of the dispersion relation in the limit of a vanishing wavevector $k \to 0$ is controlled by the Fourier transform of the power-lay-decaying interactions, and therefore it fundamentally depends on the decay exponent $\alpha$ as well as on the number of dimensions $d$. We further prove the self-consistency of the LSW approximation, showing that the LSW corrections to the mean-field solution are weak throughout the phase diagram, and even vanishing in the thermodynamic limit for $\alpha < d$. 
	

	\subsection{Model Hamiltonian}
	 The Hamiltonian of the XXZ model with LR interactions reads
	\be 
		{\cal H}_{\rm XXZ} = \sum_{i \neq j}~ \frac{J_0}{r_{ij}^{\alpha}}~ \left [ -(S_i^x S_j^x + S_i^y S_j^y) + \Delta S_i^z S_j^z \right ]
	\label{e.H_XXZ}
	\ee 
	where the indices $i,j$ run over the nodes of a $N=L^d$ hyper-cubic lattice in $d$ spatial dimensions with periodic boundary conditions (unless otherwise specified). $S_i^{\beta}$ ($\beta = x,y,z$) are quantum-spin operators attached to each node. In the following we shall specify our discussion to the case of $S=1/2$. We consider ferromagnetic interactions for $xy$ spin components, while the $z$ spin components may have either ferromagnetic ($\Delta < 0$) or antiferromagnetic ($\Delta > 0$) interactions. The coupling $J_0$ simply sets the overall energy scale, and it will be set to one in the following. 
	
The choice of $S=1/2$ and of ferromagnetic $xy$ couplings allows for a meaningful mapping of the above Hamiltonian onto that of hardcore bosons with long-range hopping and interaction
\begin{eqnarray} 
		{\cal H}_{\rm XXZ} = \sum_{i \neq j}~ \frac{J_0}{r_{ij}^{\alpha}}~ \Big [ &-& \frac{1}{2}(a_i^{\dagger} a_j  + a_j^{\dagger} a_i) \nonumber \\ 
		 &+& \Delta \left( n_i-\frac{1}{2} \right ) \left( n_j-\frac{1}{2} \right ) \Big ]
	\label{e.H_hcb}
	\end{eqnarray}
where $a_i$, $a_i^{\dagger}$, and $n_i = a_i^{\dagger} a_i$ are hardcore-boson operators (satisfying the relations $\{a_i,a_i^{\dagger}\}=1$ and  $[a_i,a_{j\neq i}^{(\dagger)}]=0$), related to the $S=1/2$ spin operators by the transformation $a_i = S_i^{-}$, $n_i - 1/2 = S_i^z$.
	
	Unless explicitely stated, we shall work with periodic boundary conditions, and choose the finite-size regularization 
	$r_{ij}^2 = \sum_{p=1}^d (\Delta r_{ij})_p^2$, with 
	\be
	(\Delta r_{ij})_p = (L/\pi) \sin(r_{ij}^{(p)} \pi / L) ~~~~~ r_{ij}^{(p)} = (\bm r_i - \bm r_j)\cdot {\bm e}_p~.
	\ee
	 The distance $r_{ij}$ is dimensionless, as it is measured in units of the lattice spacing; this choice leads to $r_{ij}^{-\alpha} \to \delta_{\langle ij \rangle}$ for $\alpha \to \infty$, where $\langle ij \rangle$ denotes a pair of nearest neighbors, namely the $\alpha \to \infty$ limit corresponds to finite-range interactions. On the opposite end, the limit $\alpha \to 0$ produces couplings with an infinite range. 
Traditionally, the distinction between short- and long-range interactions relies on the convergence properties of the sum $\sum_{j\neq i} 1/r_{ij}^{\alpha}$ in the thermodynamic limit. As the latter diverges whenever $\alpha \le d$, the separation between short- and long-range interactions is set at $\alpha = d$. 
	
	\subsection{Mean-field phase diagram}
	In $d=2$ and $d=3$, the ground-state of the Hamiltonian Eq.~\eqref{e.H_XXZ} exhibits three different phases : an Ising ferromagnetic phase (FM) with spins aligned along the $z$ direction -- corresponding to an insulating state of hardcore bosons with one particle per site; a N\'eel ordered phase with spins exhibiting a finite staggered magnetization along the $z$ axis -- corresponding to a checkerboard solid of hardcore bosons; and an XY phase where the rotational symmetry about the $z$ axis is spontaneously broken, and spins order ferromagnetically along (for instance) the $x$ axis -- corresponding to a superfluid condensate for the hardcore bosons. 
	 
	\begin{figure}
		\includegraphics[width=\linewidth]{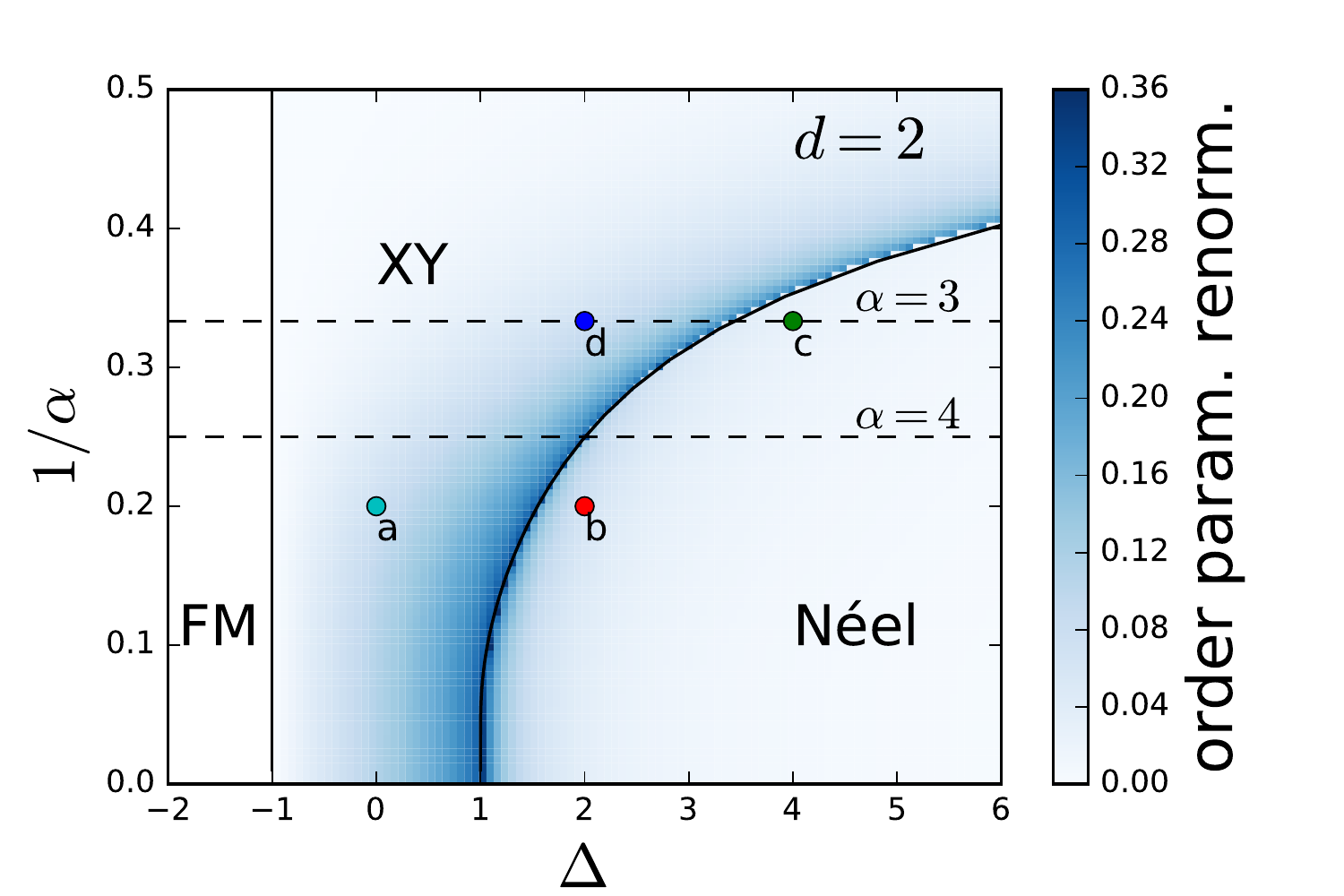}
		\caption{Phase diagram of the XXZ Hamiltonian \eqref{e.H_XXZ} in $d=2$. The false colors indicate the renormalization of the classical order parameter by quantum fluctuations, calculated on a system size $L_x = L_y = 100$. Solid lines are the mean-field prediction of Eq. \eqref{eq-E_MF}. a, b, c and d are the points were the spectrum is calculated on Fig. \ref{f.exc_spec_2d}.}
		\label{f.phase_diagram_2d}
	\end{figure}

	The approximate location of the transition lines between the different phases is predicted by a simple mean-field argument, by comparing the energy of the corresponding mean-field states : 
	\bearr
		\Psi_{\rm FM} &=& \otimes_i | \uparrow_z \rangle_i \\
		\Psi_{\textnormal{N\'eel}} &=& \otimes_{i ~{\rm even}} | \uparrow_z \rangle_i
											\otimes_{j ~{\rm odd}} | \downarrow_z \rangle_j
											\\
		\Psi_{\rm XY} &=& \otimes_i | \uparrow_x \rangle_i
	\eearr
	where even (odd) sites are located at positions ${\bm r}_i$ such that $\sum_{p=1}^d \bm r_i \cdot {\bm e}_p$ is even (odd). The corresponding energies $E = \langle \Psi \vert {\cal H}_{\rm XXZ} \vert \Psi \rangle$ are : 
	\bearr 
	\label{eq-E_MF}
		E_{\rm FM} &=& s^2 \Delta  \sum_{i\neq j} r_{ij}^{-\alpha} \nonumber \\
		E_{\textnormal{N\'eel}} &=& s^2 \Delta \sum_{i \neq j} \epsilon_i \epsilon_j r_{ij}^{-\alpha} \nonumber \\
		E_{\rm XY} &=& - s^2 \sum_{i \neq j} r_{ij}^{-\alpha}
	\eearr
	with $s=1/2$ in the present case. $\epsilon_i=1$ ($-1$) if $i$ is an even (odd) site.
	For $\Delta < -1$, the Ising ferromagnet (FM) has the lowest energy, and in fact it corresponds to the exact ground state: given that this state does not possess quantum correlations nor entanglement, we shall not discuss it any further, and we will restrict our attention to the case $\Delta > -1$. In the latter range, the condition $E_{\textnormal{N\'eel}}=E_{\rm XY}$ sets the transition line at 
  \be
	\Delta_c(\alpha) =  - \frac{\sum_{i \neq j} r_{ij}^{-\alpha} }{ \sum_{i \neq j} \epsilon_{i} \epsilon_j r_{ij}^{-\alpha} }
	\label{e.Deltac}
	\ee  
	Notice that, since $r_{ij}^{-\alpha}$ decays with distance between $i$ and $j$, the denominator is always negative, so that the above ratio is positive. For $\Delta > \Delta_c(\alpha)$, the system displays N\'eel order, while for $\Delta < \Delta_c(\alpha)$, the in-plane (XY) ferromagnetic order is favored. For $\alpha \to \infty$, the interactions are restricted to nearest neighbor, and one recovers the known result $\Delta_c(\infty) = 1$. Due to the frustration of the interaction among $z$ spin components inherent to the long-range nature of the couplings, we expect that when $\alpha$ decreases, a larger value of $\Delta$ is necessary to stabilize the N\'eel order. And indeed, one can predict that at $\alpha < d$, the XY order is always stabler than the N\'eel order, since $\sum_{i \neq j} \epsilon_{i} \epsilon_j r_{ij}^{-\alpha}$ is always finite, while $\sum_{i \neq j} r_{ij}^{-\alpha}$ diverges. One thus obtains the phase diagram shown on Fig. \ref{f.phase_diagram_2d}. When including harmonic quantum corrections to the mean-field solution (as discussed in the following) the phase diagram preserves its topology in $d=2$ and 3. In $d=1$, on the other hand, the XY phase is destabilized for $\alpha > 3$, in compliance with Mermin-Wagner theorem \cite{Auerbach-book}, as further discussed in Sec.~\ref{s_stability} (the correct phase diagram for $d=1$ including quantum corrections to mean-field theory is to be found in Fig.~\ref{f.phase_diagram_1d}). 
	 

	\subsection{Spin waves and excitation spectrum}
	\label{s.spectrum} 
	In this section we study the low energy properties of the XXZ Hamiltonian through linear spin-wave theory, representing a semi-classical expansion around the mean-field ground-state \cite{colettaetal2012}.
	\\
	 
	 \subsubsection{XY phase}
	 The mean-field ground-state in the XY phase is a perfect ferromagnet with all spins aligned along \emph{e.g.} the $x$ axis. We then introduce Holstein-Primakoff (HP) boson operators $b_i$ which describe small deviation with respect to this perfect ferromagnet (namely the spin waves) \cite{holstein_primakoff, colettaetal2012}:
	\bearr
		S_i^x & = &  \frac{1}{2} - b_i^\dagger b_i \nonumber \\
		S_i^y & = & \frac{1}{2i}(b_i - b_i^\dagger) + O(b_i^3) \nonumber \\
		S_i^z & = & -\frac{1}{2}(b_i + b_i^\dagger) + O(b_i^3)
	\label{e.HP-transform_XY}
	\eearr
	The LSW approximation consists in discarding all terms beyond quadratic in the HP transformation as well as in the resulting Hamiltonian. Its validity relies on the assumption that the quantum fluctuations of the spins are not strong enough to destroy the classical order. Technically, this hypothesis requires the populations of the HP bosons $\langle b_i^\dagger b_i \rangle$ to remain significantly smaller than 1/2 (or $s$ for a generic spin length $s$). A systematic comparison of some predictions of the LSW expansion with exact Monte-Carlo calculations in $d=2$ has been presented in Ref.~\cite{colettaetal2012}, in the $\alpha = \infty$ (nearest-neighbor) case and at $\Delta=0$, demonstrating a remarkable accuracy for the LSW results. The accuracy of LSW predictions can only be expected to improve upon lowering the $\alpha$ exponent, as long as this does not entail crossing the phase transition separating the XY phase from the N\'eel phase, as discussed in the previous section. If instead one considers values of $\alpha$ and $\Delta$ close to the transition line in Fig.~\ref{f.phase_diagram_2d} one may expect the quality of LSW theory to worsen: nonetheless, as discussed in Sec.~\ref{s_stability} the first-order nature of the XY-N\'eel transition guarantees that our LSW treatment remains justified. 
	 
Taking advantage of the translational invariance, the Hamiltonian is conveniently expressed in momentum space. Introducing $b_{\bm k} = N^{-1/2} \sum_j e^{-i {\bm k} \cdot {\bm r}_j} b_j$, and expanding the Hamiltonian up to second order in the $b_{\bm k}$ operators, one obtains 
	\be
		{\cal H}^{(2)} = \frac{1}{2} \sum_{\bm k} 
				\begin{pmatrix}
					b_{\bm k}^\dagger & b_{\bm k}
				\end{pmatrix} 
	   			\begin{pmatrix}
	   				A_{\bm k} & B_{\bm k} \\ 
	   				B_{\bm k} & A_{\bm k}
	   			\end{pmatrix}
	   			\begin{pmatrix}
	   				b_{\bm k} \\ b_{\bm k}^\dagger
	   			\end{pmatrix}
	\label{e.H_quadra}	  
	  \ee
	  where a constant term has been dropped. We have introduced the coefficients 
	  \begin{equation}
	  A_{\bm k} = \gamma_0 + (\Delta -1) \gamma_{\bm k} / 2~~~~ B_{\bm k} = (\Delta + 1) \gamma_{\bm k} / 2
	  \label{e.AkBkXY}
	  \end{equation}
	  and the fundamental geometric coefficient 
	  \begin{equation}
	  \gamma_{\bm k} = \sum_{\bm r \neq 0} \frac{e^{i {\bm k}\cdot \bm r}}{r^{\alpha}}
	  \end{equation} 
	  corresponding to the Fourier transform of the coupling matrix.
	  
	  The quadratic Hamiltonian ${\cal H}^{(2)}$ is diagonalized by a Bogoliubov transformation \cite{blaizot-ripka} $b_{\bm k} = u_{\bm k} \beta_{\bm k} - v_{\bm k} \beta_{-\bm k}^\dagger$, with coefficients $u_{\bm k} = (1/\sqrt{2})(A_{\bm k} / \sqrt{A_{\bm k}^2 - B_{\bm k}^2} + 1)^{1/2}$ and $v_{\bm k} = \sqrt{u_{\bm k}^2 - 1}$, to lead to the form: 
	  \be
	  		{\cal H}^{(2)}_{\rm XY} = \sum_{k\neq 0} \gamma_0 \sqrt{1 - \frac{\gamma_{\bm k}}{\gamma_0}}  \sqrt{1 + \Delta \frac{\gamma_{\bm k}}{\gamma_0}} \beta_{\bm k}^\dagger \beta_{\bm k} + {\cal H}_{k=0}~.
	\label{e.H_quadra_final_XY}	  
	  \ee
	  
\begin{figure}
			\includegraphics[width=\linewidth]{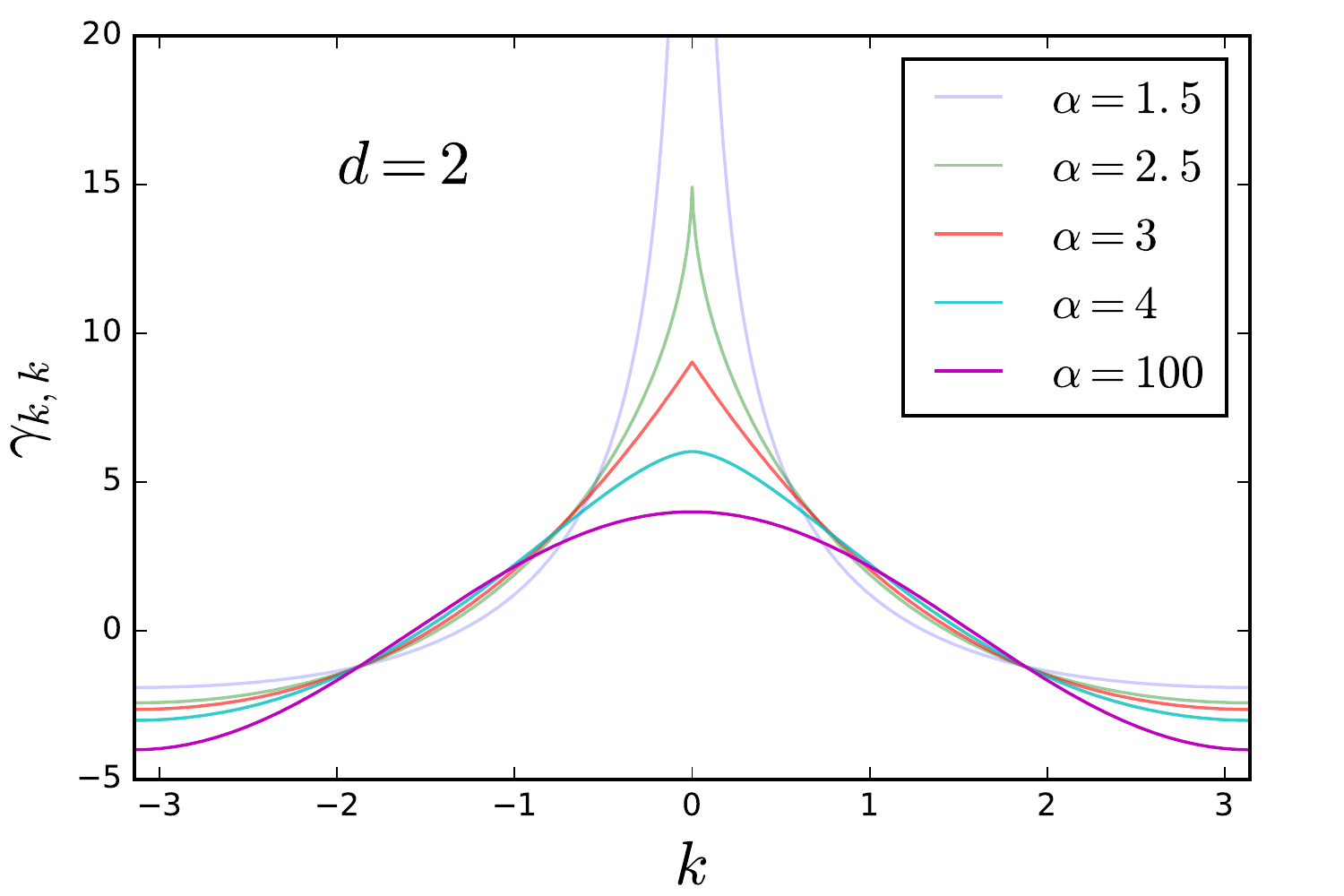}	  		
		\caption{Fourier transform $\gamma_{\bm k}$ of the interaction $r_{ij}^{-\alpha} = 1/| r_i - r_j |^\alpha$, in dimension $d=2$. $\gamma_{\bm k}$ is plotted along the $(k,k)$ diagonal of the Brillouin zone. For $\alpha < d$, $\gamma_{\bm k}$ diverges as $k^{\alpha - d}$ at small $k$. For $d < \alpha$, $\gamma_{\bm k} - \gamma_0 \sim k^{\min(2, \alpha - d)}$ (see text).}
		\label{f.gamma_k_2d}
	\end{figure}

	   In fact, the Bogoliubov transformation only applies to $k \neq 0$, while the $k=0$ sector deserves a special treatment.
	  When $\alpha > d$, $\gamma_0 = \sum_{r\neq 0} 1/r^{\alpha}$ is non-divergent, and therefore the $k=0$ sector provides a microscopic contribution to the Hamiltonian and to the thermodynamic properties \footnote{Note that $\gamma_{\bm k}$ converges for any nonzero $k$ and any $\alpha > 0$.}. For $\alpha > d$, ${\cal H}_{k=0}$ can then be safely ignored in the calculations when taking the thermodynamic limit. The $\beta_{\bm k}$ operators describe quasi-particles which, at the LSW level of approximation, correspond to exact eigenmodes of the many-body Hamiltonian. They represent collective fluctuations of the spins above the ground-state, with a dispersion relation 
	  \be
	  E_{\bm k} = \sqrt{A_{\bm k}^2 - B_{\bm k}^2} = \gamma_0\sqrt{1 - \gamma_{\bm k}/\gamma_0} \sqrt{1 + \Delta \gamma_{\bm k}/\gamma_0}~.
	  \ee 
	   The LSW approximation is dynamically stable if $E_{\bm k}$ is real for any $\bm k$. Since $\gamma_{\bm k} \le \gamma_0$, the term  $\sqrt{1 - \gamma_{\bm k}/\gamma_0}$ is not problematic. 
	  As $\gamma_{\bm k}$ is maximally negative for $\bm K=(\pi, \pi, \dots)$ (see Fig. \ref{f.gamma_k_2d}), we must require $\Delta < - \gamma_0 / \gamma_{\bm K}$, coinciding with the condition $\Delta < \Delta_c(\alpha)$, Eq.~\eqref{e.Deltac}, for the mean-field stability of the XY phase; this condition must be supplemented with $\Delta > -1$ to avoid the instability towards the Ising ferromagnetic phase. The resulting spectrum is gapless at $k=0$, corresponding to the Goldstone mode associated with the broken $U(1)$ rotational symmetry around the $z$ axis. An important qualitative difference with the short-range regime is that the usual linear dispersion $\omega \sim k$ for this Goldstone mode, is altered for $d< \alpha < d+2$ into $\omega \sim k^z$, with $z = (\alpha - d)/2$, as we shall discuss below. For $\alpha < d$, on the other hand, one observes that $\gamma_0 \to \infty$, while $\gamma_{\bm k}$ remains convergent. As a consequence, the excitation spectrum (rescaled in units of $\gamma_0$) becomes \emph{dispersionless} in the thermodynamic limit. We further discuss the case $\alpha < d$ in Section \ref{sec-infinite_range}. 
	  
	  \textit{Dynamical exponent.---}
	   The geometric coefficient $\gamma_{\bm k}$ controls the low energy dispersion relation $E_{\bm k}$ (see Eq. \eqref{e.H_quadra_final_XY} and Eq. \eqref{e.H_quadra_final_Neel}), and the large distance decay of the spin correlations is also directly controlled by the small $k$ behavior of $\gamma_{\bm k}$ (see Appendix \ref{app-correl}). In order to obtain the correct scaling regimes, independently of the details of the lattice (as expected in the small-$k$ limit), and to recover simultaneously the correct dispersion relation in the limit $\alpha \to \infty$, we found convenient to treat exactly the nearest-neighbor contribution to $\gamma_{\bm k}$, and to approximate the rest of the sum by an integral. We thus have : 
	  		\be
	  			\gamma_{\bm k} = \gamma_{\bm k}^{\rm (nn)} +  \int_{\rho > 1} d^d \rho ~ \frac{e^{i a {\bm k}\cdot{\bm \rho} }}{\rho^{\alpha }}
				\label{e.gamma}
	  		\ee
	  		where ${\bm \rho} = {\bm r}/a$, $a$ is the lattice constant, and $\gamma_{\bm k}^{\rm (nn)}$ ($= 2\sum_{i=1}^d \cos(k_i a)$ on the cubic lattice under the present investigation) is the value of $\gamma_{\bm k}$ for $\alpha \to \infty$, \textit{i.e.} for nearest-neighbor interactions only. For any $\alpha> d$, $\gamma_0$ is expressed via a convergent integral. In Appendix \ref{app-gamma_k}, we show that $\gamma_{\bm k}$ has the following behavior at small $k$ : 
	  		\bearr
	  			\gamma_0 - \gamma_{\bm k} & \sim & k^{\alpha - d} ~~~ \textnormal{(for $\alpha  < d + 2$)} \nonumber \\ 
	  			\gamma_0 - \gamma_{\bm k}  &\sim & k^2 ~~~ \textnormal{(for $\alpha > d + 2$)} 
	  		\label{e.scaling-Jk}
	  		\eearr	
As a consequence $\sqrt{\gamma_0 - \gamma_{\bm k}}$ develops a cusp around $k=0$ for $d < \alpha < d+2$, turning into a divergence for $\alpha < d$ (see Fig. \ref{f.gamma_k_2d}).	
				
	  		Interestingly, although the calculation of the integral in Eq.~\eqref{e.gamma} requires a different treatment for each value of $d$, the low-$k$ dispersion shows a clear change at the simple $d$-dependent value of the $\alpha$ exponent, $\alpha = d + 2$. When $\alpha > d + 2$, the qualitative behavior of $\gamma_0 - \gamma_{\bm k}$ is the same as for the short range limit $\alpha  \to \infty$, with a prefactor of the $k^2$ scaling that depends on $\alpha $ and on the details of the lattice, and which diverges at $\alpha  = d + 2$. 
	  		On Fig. \ref{f.gamma_k_2d}, we have plotted $\gamma_{\bm k}$ in $d=2$ along the $(k,k)$ diagonal of the Brillouin zone.

\textit{Excitation spectrum and sound velocity .---}
		Given the expressions of the excitation spectrum $E_{\bm k}$ in the XY phase Eq. \eqref{e.H_quadra_final_XY}, and given the discussion of the previous paragraph (see Eq. \eqref{e.scaling-Jk}), the small $k$ behavior of the excitation spectrum is straightforwardly derived.

	
		\begin{figure}
		\includegraphics[width=\linewidth]{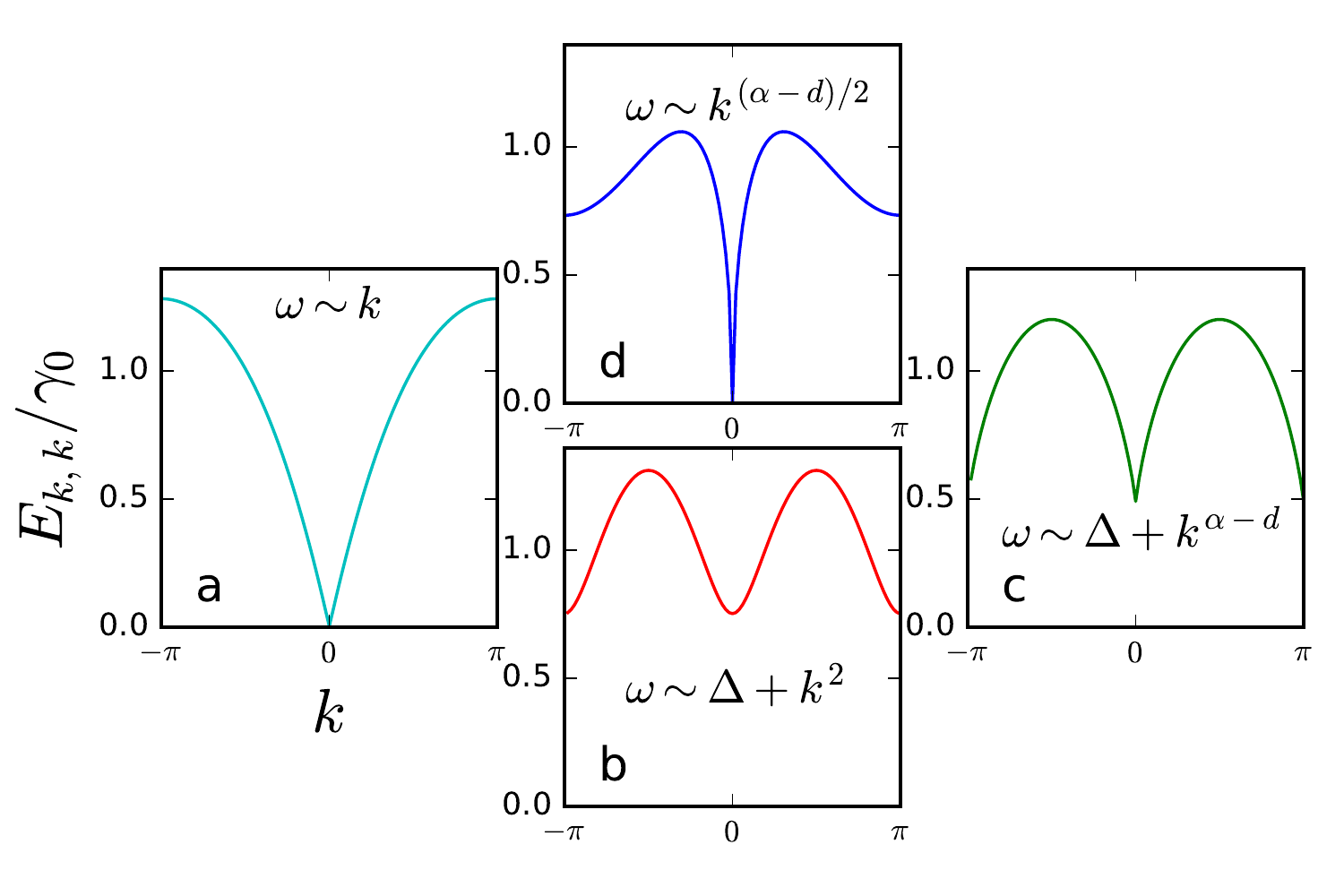}
		\caption{Excitation spectrum along $k_x = k_y$ in dimension $d=2$. (a) $\Delta = 0$ and $\alpha=5$ (XY phase); (b) $\Delta = 2$ and $\alpha = 5$ (N\'eel phase); (c) $\Delta = 4$ and $\alpha = 3$ (N\'eel phase); (d) $\Delta = 2$ and $\alpha = 3$ (XY phase). The a,b,c and d points are indicated on the phase diagram, Fig. \ref{f.phase_diagram_2d}.}
		\label{f.exc_spec_2d}
	\end{figure}
	
	The result is summarized on Table \ref{table_summary}, and illustrated on Fig. \ref{f.exc_spec_2d} at various representative points of the phase diagram. The spectrum is gapless in the whole XY phase, corresponding to the Goldstone mode of the broken U(1) rotational symmetry. Remarkably, the linear dispersion relation of this Goldstone mode, well known in the short range regime [Fig. \ref{f.exc_spec_2d}(a)] is recovered only when $\alpha > d+2$, with an $\alpha$- and $\Delta$-dependent sound velocity $\partial E_{\bm k} / \partial k |_{k=0}$; while for $d< \alpha < d + 2$, the dispersion relation behaves as $k^z$, with $z=(\alpha - d) / 2 < 1$, and the group velocity diverges as $k^{z - 1}$ at small $k$ [Fig. \ref{f.exc_spec_2d}(d)].

	  \subsubsection{N\'eel phase}
	  In the N\'eel phase, the Holstein-Primakoff bosons represent small deviations around the reference mean-field state $\Psi_{\textnormal{N\'eel}}$, and they are introduced via the following transformation: 
	\bearr
		S_i^z & = &  \epsilon_i \left(\frac{1}{2} - b_i^\dagger b_i \right) \nonumber \\
		S_i^y & = & \epsilon_i ~\frac{1}{2i}(b_i - b_i^\dagger) + O(b_i^3) \nonumber \\
		S_i^x & = & \frac{1}{2}(b_i + b_i^\dagger) + O(b_i^3)
	\label{e.HP-transform_Neel}
	\eearr
	with $\epsilon_i=1$ for even sites and $\epsilon_i=-1$ for odd sites. The reduction to an effective quadratic Hamiltonian goes along the same line as for the XY ferromagnet. The $A_{\bm k}, B_{\bm k}$ coefficients take the form (with again $\bm K=(\pi, \pi, \dots)$)
	\begin{eqnarray}
	A_{\bm k} &=& -\Delta \gamma_{\bm K} - (\gamma_{\bm k} + \gamma_{{\bm k}-{\bm K}}) / 2 \nonumber \\ 
	B_{\bm k} &=& - (\gamma_{\bm k} - \gamma_{{\bm k}-{\bm K}}) / 2~
	\label{e.AkBkNeel}
	\end{eqnarray}
	resulting in the following quadratic Hamiltonian
		\be
	  		{\cal H}^{(2)}_{\textnormal{N\'eel}} = \sum_{k} \sqrt{(\Delta \gamma_{\bm K} + \gamma_{\bm k}) (\Delta \gamma_{\bm K} + \gamma_{{\bm k}-{\bm K}}) } \beta_{\bm k}^\dagger \beta_{\bm k} ~.
	\label{e.H_quadra_final_Neel}	  
	  \ee
	  The spectrum is gapped and twofold degenerate (because it is folded onto the smaller magnetic Brillouin zone, which is half the geometric one) with $E_{\bm k} = E_{{\bm k}-{\bm K}}$ (see Fig. \ref{f.exc_spec_2d}). The stability requirement that the spectrum be real is equivalent to the condition $\Delta > - \gamma_0 / \gamma_{\bm K}$, which is again the same criterion defining the phase boundary between the N\'eel phase and the XY ferromagnetic phase. Near $k=0$, the spectrum behaves as $E_{\bm k} \approx \Delta_g + c k^{z}$, where $\Delta_g = \gamma_{\bm K} \sqrt{\Delta + 1} \sqrt{\Delta + \gamma_0 / \gamma_{\bm K}}$, $c$ is some ($\alpha-$ and $\Delta-$dependent) constant and $z = \min(\alpha-d, 2)$. The dispersion relation expected in the short-range regime is therefore recovered when $\alpha > d + 2$ [Fig. \ref{f.exc_spec_2d}(b)]. Moreover, for $d< \alpha < d + 1$, the maximal group velocity (moving from a finite $k$ towards $k=0$) diverges as $v_G^{\max} =  \max_{\bm k} | 
	  \bm \nabla_{\bm k} E_{\bm k}| \sim k^{\alpha - d -1}$ [Fig. \ref{f.exc_spec_2d}(c)].

	\subsection{Stability of the spin-wave approximation}
	\label{s_stability}
		In this section, we briefly discuss the self-consistency of the LSW approximation in the different regimes. Self-consistency requires that the modification of the order parameter $m_{\rm SW}$, due to the non-zero population of Holstein-Primakoff bosons, remains small with respect to its classical value $m_{\rm cl} = 1/2$. In the XY phase, the order parameter $m$ is the average magnetization along $x$, while it is the staggered magnetization in the N\'eel phase. According to the Holstein-Primakoff transformations Eq. \eqref{e.HP-transform_XY} and \eqref{e.HP-transform_Neel}, one thus has to evaluate $(m_{\rm cl} - m_{\rm SW})/m_{\rm cl} = 2\langle b_i^\dagger b_i \rangle = (2/N) \sum_{\bm k} n_{\bm k}$, with $n_{\bm k} = \langle b_{\bm k}^\dagger b_{\bm k} \rangle = v_{\bm k}^2$. This integral is always finite in the gapped N\'eel phase, but could possibly diverge in the XY gapless phase due to the small $k$ behavior of $v_{\bm k}^2 \sim 1/E_{\bm k} \sim k^{-z}$: this happens if $z \ge d$. Given that  $z = 1$  for $\alpha \ge d+2$ and $z<1$ otherwise, the only true instance of instability of the LSW approximation is found in $d=1$ for $\alpha > 3$ in the XY phase, featuring a logarithmic divergence in the system size. 
		The strongest renormalization of the order parameter due to quantum fluctuations is found for $\alpha \to \infty$ and $\Delta = 1$, namely at the Heisenberg point of the XXZ model with nearest-neighbor interactions. Therefore in $d=2$ and $d=3$ the renormalization of the order parameter throughout the $\Delta$-$\alpha$ phase diagram is upped bounded by that of the nearest-neighbor Heisenberg model, namely 40\% on the $d=2$ square lattice and 16\% on the $d=3$ cubic lattice \cite{Mattis-book}. This means in particular that the N\'eel-to-XY quantum phase transition at $\alpha < \infty$ does not possess stronger quantum renormalization than those at the Heisenberg point, and therefore it is still quantitatively described by LSW theory. This is intrinsically due to the \emph{first-order} nature of the phase transition, which implies that true quantum critical fluctuations and entanglement do not develop.   
		
\begin{figure}
		\includegraphics[width=\linewidth]{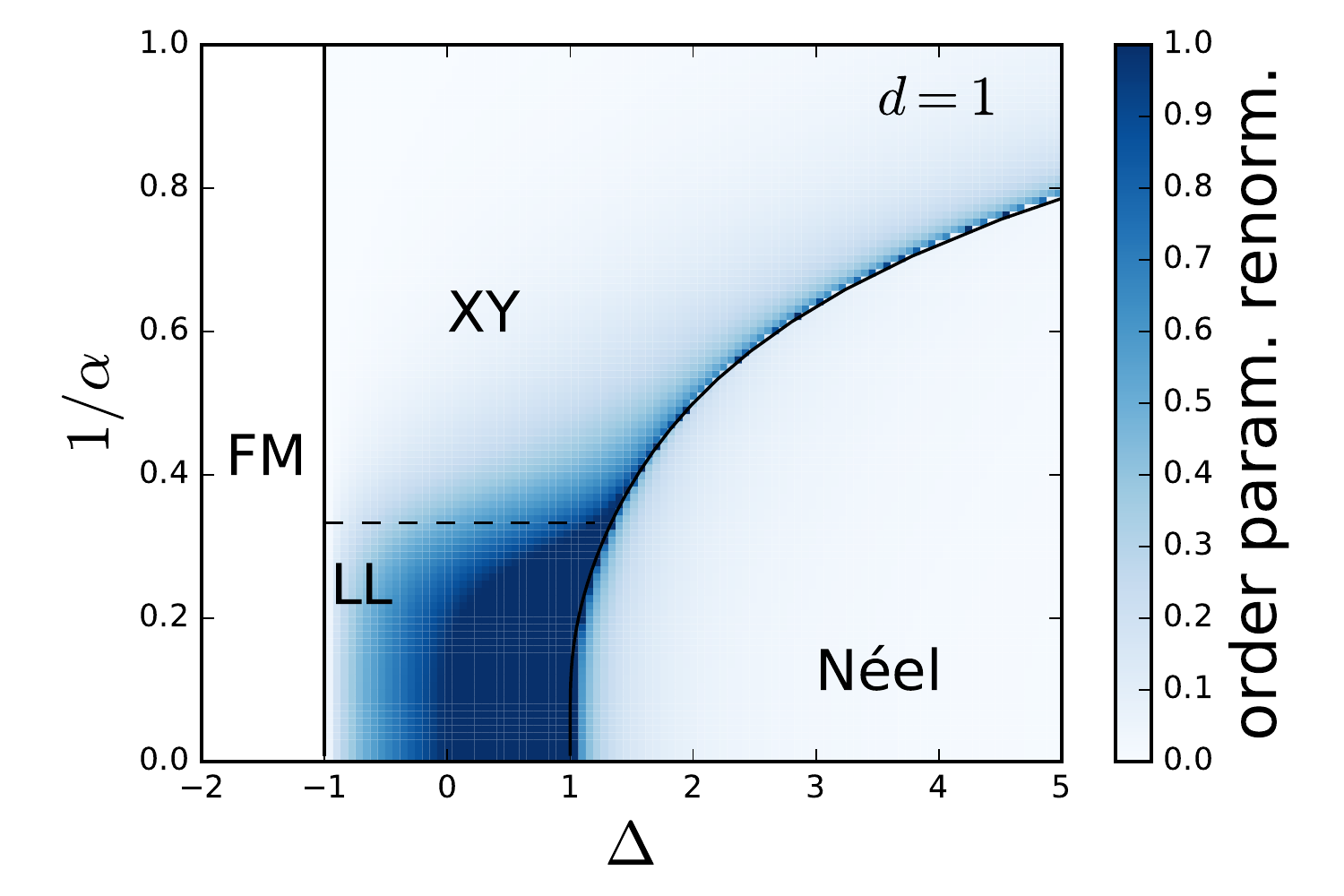}
		\caption{Phase diagram of the XXZ Hamiltonian \eqref{e.H_XXZ} in $d=1$. The false colors indicate the renormalization of the classical order parameter by quantum fluctuations as calculated on a system of size $L = 1000$. Solid line is the mean-field prediction of Eq. \eqref{eq-E_MF}, while the dashed line marks the breakdown of LSW theory at $\alpha = 3$, signaling the XY-LL transition. See \cite{maghrebietal2015} for a more complete study in $d=1$.
		}
				\label{f.phase_diagram_1d}
	\end{figure}

		In $d=1$, the renormalization of the classical order diverges in the thermodynamic limit when $\alpha \to 3^-$ -- a finite-size system the calculation still delivers a finite renormalization, \emph{e.g.} for $\alpha = 3$ and $L = 10^5$ the renormalization reaches $86 \%$. 
	The breakdown of LSW theory for $\alpha \ge \alpha_c = 3$ signals a phase transition, absent at the mean-field level, between a long-range-ordered XY ferromagnet (or a superfluid condensate) for $\alpha < \alpha_c$ and a quasi-long-range-ordered Luttinger liquid (LL) phase \cite{Giamarchi-book} for $\alpha>\alpha_c$. The phase diagram in $d=1$ is reported in Fig. \ref{f.phase_diagram_1d}. This quite unusual phase transition, specific to $d=1$, has been studied in more details in \cite{maghrebietal2015, gongetal2016} beyond the harmonic approximation, while LSW theory can only indicate the existence of such a transition, but it cannot quantitatively describe it. In particular, as shown in Ref.~\cite{maghrebietal2015} the phase transition occurs at a $\Delta$-dependent value $\alpha_c(\Delta) < 3$. In view of the limitations of LSW results in this regime, in Sec.~\ref{sec-TOS_1d} we shall complement them with fully quantitative ones based on the DMRG approach, in order to investigate the evolution of entanglement properties across the XY-LL transition. 
	  
	\subsection{Exact ground state for $\alpha < d$}
	\label{sec-infinite_range}
	  	When $0 < \alpha < d$, LSW theory predicts that classical order is not renormalized by quantum fluctuations, irrespective of the dimension $d$ and of the precise value of $\alpha$. Indeed, in this long-range regime  -- which for $\Delta > -1$ always falls into the XY phase -- $\gamma_0$ diverges, while $\gamma_{\bm k}$ goes to a constant for any fixed nonzero $k$ in the thermodynamic limit. This implies that $A_{\bm k} / B_{\bm k} \to 0$ at any $\bm k \neq 0$, so that $u_{\bm k} \to 1$ and $v_{\bm k} \to 0$ in the thermodynamic limit. In other words, the spin waves $b_{\bm k}^\dagger |0\rangle$ become exact eigenstates of the Hamiltonian, whose ground state \emph{is} the mean-field trial state. One is thus left with an effective dynamics for the ground state which occurs only in the $k=0$ sector (the collective spin ${\bm S}_{\rm tot} = \sum_i {\bm S}_i$). One might question the reliability of the LSW approach to draw a definite conclusion on the nature of the ground state. However, we can show that the prediction of LSW theory is essentially exact: the XY ferromagnetic state, while not being the exact ground state on a finite size system, has a vanishingly small energy density above the ground state in the thermodynamic limit. In order to do so, we first rewrite the Hamiltonian as : 
	  	\be 
	  		H_{\rm XXZ} = \sum_{\bm k} \gamma_{\bm k} \left ( -S_{\bm k}^x S_{-\bm k}^x - S_{\bm k}^y S_{-\bm k}^y + \Delta S_{\bm k}^z S_{-\bm k}^z \right )
	  	\ee
	  	where we have introduced $S_{\bm k}^\beta = N^{-1/2} \sum_i e^{i{\bm k}\cdot{\bm r}_i} S_i^\beta$. Since $\gamma_{k\neq 0} / \gamma_0 \to 0$ in the thermodynamic limit, we may keep only the $k=0$ sector, which reproduces the so-called Lipkin-Meshkov-Glick model \cite{Latorreetal2005}:
	  	\be 
	  	H_{\rm XXZ} / \gamma_0  \approx \frac{1}{N} \left [ -({\bm S}_{\rm tot})^2 + (1+\Delta) (S_{\rm tot}^z)^2 \right ]~.
	  	\ee
	  The latter model is readily diagonalized by the total-spin basis $|S,M\rangle$ to give: 
	  	\be 
	  		H_{\rm XXZ}  | S, M \rangle = \frac{\gamma_0}{N} \left [ -S(S+1) + (1+\Delta) M^2 \right ] | S, M \rangle~.
	  	\ee
	  	As long as $\Delta > -1$, the ground-state is 
	  	\be
	  		| \Psi_0 \rangle = |S = N/2,~ M = 0 \rangle 
	  	\ee
	  	with an energy $E_0 = -N/4 - 1/2$ (since $\gamma_0$ diverges for $N\to\infty$, we measure energy in units of $\gamma_0$, so that the thermodynamic limit is well defined).
	  	The state $| \Psi_0 \rangle$ -- often referred to in the atomic-physics literature as the \emph{spin-squeezed} state -- is obtained from the application of $(\sum_i \sigma_i^+)^{N/2}$ to the state with all spins down, where $\sigma_i^+$ flips the spin at position $i$ if it is a spin down, and gives zero if applied to a spin up. Being the symmetric, equal weight superposition of all possible spin configurations with $N/2$ spins up and $N/2$ spins down, it contains no further correlation between the spins beyond the global constraint that the total spin along $z$ be zero. And, in the thermodynamic limit, the mean-field XY ferromagnet (which does not possess any form of correlation), has a vanishingly small energy density compared to the squeezed state:
	  	\be 
	  	  (E_{\rm XY} - E_0) / N = 1 / 2N \to 0
	  	 \ee
	  	 and it has essentially the same properties as the true ground state. Another way to understand this result, following the early works of Anderson \cite{Anderson1952,Anderson-book}, is to realize that the mean-field XY ferromagnet also lies in the $S=N/2$ manifold, but it is a superposition of many states with different values of $M$. Those states differ in energy with respect to the ground state by $E(N/2, M) - E_0 = (\Delta + 1) (M^2 / N)$ which scales to 0 as $1/N$ --- building up the famous ``tower of states''. Any superposition of states in the $S=N/2$ sector sufficiently peaked around $M=0$ thus becomes degenerate with the true ground state in the thermodynamic limit. This latter point of view only partly applies to the XY ferromagnet, since the distribution of $M$ has, in that case, a width of order $O(\sqrt{N})$, and hence an excitation energy of order ${\cal O}(1)$, as we have shown above.
	  	If the behavior on finite-size systems can strongly deviate from the mean-field one, one can nevertheless conclude that any correlation effect for $\alpha < d$ is a finite size effect. The entanglement properties in the long-range regime will be further discussed in Sec.~\ref{s_EELR}

	\section{Structure of quantum correlations and fluctuations in the ground state}
	\label{s_quantum_correlations}
	
		\subsection{Decay of correlation functions}

		\begin{table}
		\begin{tabular}{|c | c c |}
		\toprule
		 Correlation functions &  \hspace{20pt} XY    \hspace{20pt}&   \hspace{20pt} N\'eel    \hspace{20pt}  \\
		 $\langle S_i^\nu S_j^\nu \rangle - \langle S_i^\nu \rangle \langle S_j^\nu \rangle$  & \hspace{10pt} $ \sim 1 / r^{\eta_\nu}$ & $ \sim e^{-r/\xi} + 1 / r^{\eta_\nu}$ \\
		 \hline
		& & \\
		  $\eta_x$ & (long.) $2 (d-z)$ & (trans.) $\alpha$ \\
		  $\eta_y$ & (trans.) $d-z$ & (trans.) $\alpha$ \\
		  $\eta_z$ & (trans.) $d+z$ & (long.) $2\alpha$  \\
		  \hline
		
		\end{tabular}
		\caption{ \label{table_correl_functions} Decay exponent of the correlation functions for $\alpha > d$. ``Long.'' and ``trans.'' stand for longitudinal and transerve (with respect to the direction of the order parameter). In the XY phase, $z=\min[1, (\alpha-d)/2]$ is the dynamical exponent of Table \ref{table_summary}. In the N\'eel phase, the gap induces a short distance exponential decay, while the algebraically decaying interaction as $1/r^\alpha$ gives a long distance algebraic tail for arbitrary $\alpha$. The correlation function $\langle  S_i^z S_j^z \rangle -  \langle  S_i^z \rangle \langle  S_j^z \rangle < 0$ for $i \neq j$ in the XY phase, and shows a staggered pattern in the N\'eel phase.}
		\end{table}
			
		In this Section, we discuss the LSW predictions for the large-distance decay of the spin correlations in the ground-state. We focus on $\alpha>d$, since all correlations vanish in the thermodynamic limit for $\alpha<d$ -- they just stem from the conservation of $S^z=\sum_i S^z_i$ (see  Section \ref{sec-infinite_range}). Our observations are summarized in Table \ref{table_correl_functions}, and the calculations leading to the various spin correlations are detailed in Appendix \ref{app-correl}.

		\subsubsection{XY phase}
		
\begin{figure}
		\includegraphics[width=\linewidth]{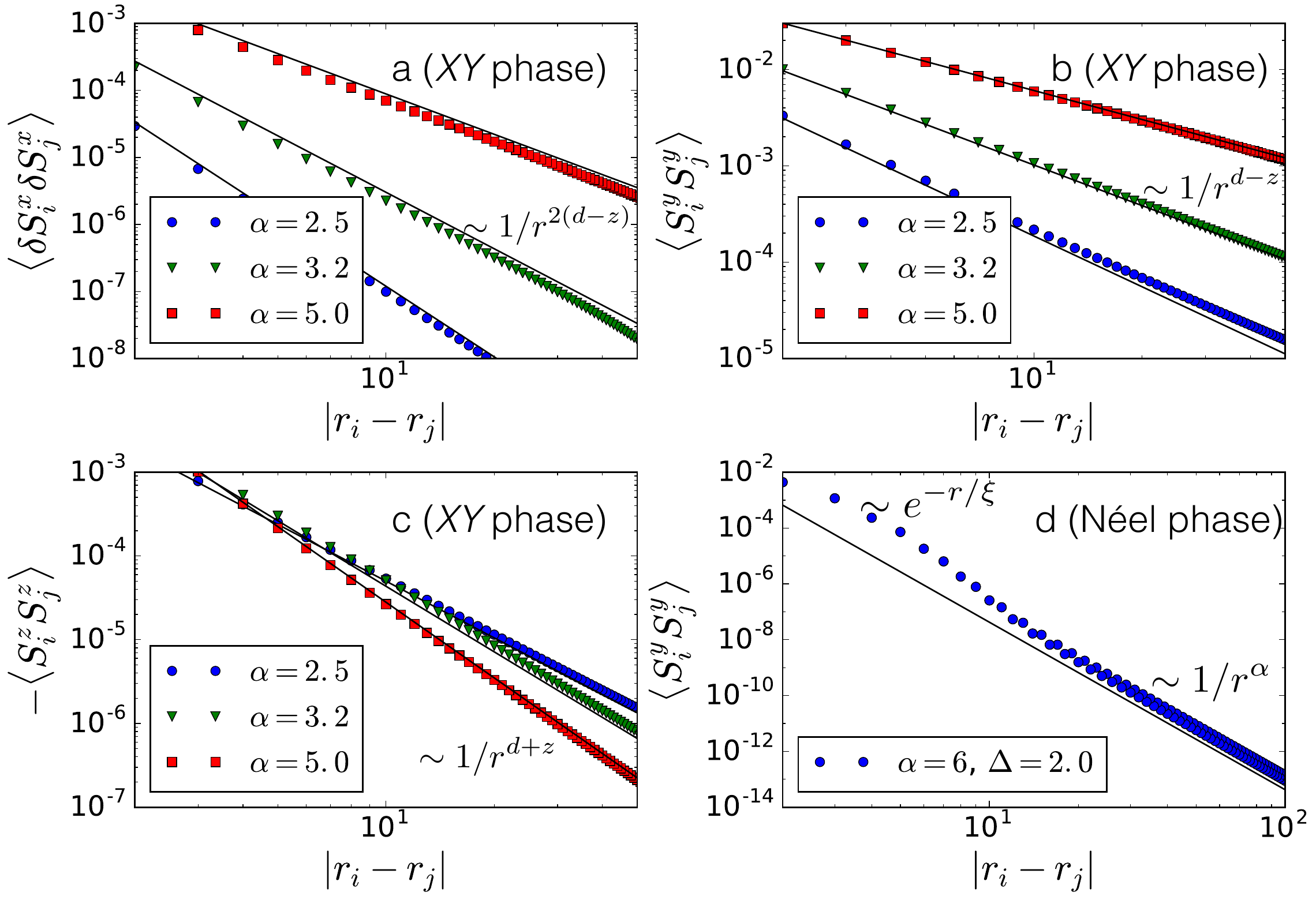}
		\caption{Spin correlation in the XY phase (a, b, c) and $S^y S^y$ correlations in the N\'eel phase (d). We chose $\Delta=0$ in the XY phase, and $L_x = 5000$, $L_y = 1000$. Deviations from the theoretical scaling are finite size effects, mainly due to the fact that the $k=0$ mode has been removed from the calculations. In the N\'eel phase, $S^z S^z$ correlations (not showed) exhibit a staggered pattern. Black solid lines indicate the power-law decays in accordance with the predictions of Table  \ref{table_correl_functions} - when possible, they are offset with respect to the LSW data for better readability of the figure.}
		\label{f.C_2D}
	\end{figure}

		 In the XY phase, the rotational symmetry about the $z$ axis is spontaneously broken. As a result, the spin correlations in the $xy$ plane are not isotropic : the $S^yS^y$ spin correlations (transverse to the order parameter), do not possess the same decay as the $S^x S^x$ spin correlations (longitudinal to the order parameter). The $S^zS^z$ spin correlations exhibit a third distinct decay behavior. Fig. \ref{f.C_2D} shows the various spin correlations in the XY phase for $d=2$.

 LSW predicts the following behaviors:
			\bearr
				\langle S_i^x S_j^x \rangle - \langle S_i^x \rangle \langle S_j^x \rangle &\sim & 1 / r^{2(d-z)} \\
				\langle S_i^y S_j^y \rangle &\sim & 1 / r^{d-z} \\
				\langle S_i^z S_j^z \rangle &\sim & 1 / r^{d+z}
			\eearr
			An analytical understanding for these decay exponents can be obtained from the small-$k$ behavior of the structure factor $S_{\bm k}^{\beta\beta} = N^{-1} \sum_{i,j} e^{i\bm k \cdot (\bm r_j - \bm r_i)} \langle S_i^\beta S_j^\beta \rangle$ ($\beta = x,y,z$). In Appendix \ref{app-correl}, we show indeed that $S_{\bm k}^{zz} \sim k^z$ while $S_{\bm k}^{yy} \sim k^{-z}$; then the decay exponent of the correlation function can be related to that of the small-$k$ of the corresponding structure factor via the calculation of Appendix \ref{app-gamma_k} , assuming algebraically decaying correlations. Moreover the fact that the decay exponent of the $S^xS^x$	 correlation function is twice the one of the 	$S^yS^y$ correlation function can be understood in that $S^xS^x$ is quartic in the HP bosons, while $S^yS^y$ is quadratic (see Appendix \ref{app-correl}). 
As the large-distance decay of the spin-spin correlations is controlled by the $\alpha$- and $d$-dependent $z$ exponent, we can conclude that the same decay exhibited for finite-range interactions ($\alpha = \infty$) is recovered over the whole short-range regime $\alpha > d+2$; while in the medium-range regime ($d < \alpha < d+2$) the decay exponents continuously depend on $\alpha$ exhibiting a broad variety of different correlation regimes. Note that the exact ground state state on a finite-size system does not break the U(1) symmetry; and the LSW prediction for the $\eta_y$ exponent is the one which reproduces the power-law decay of correlations in the symmetric case.

			\subsubsection{N\'eel phase}
			
	In the N\'eel phase, the rotational symmetry about the $z$ axis is preserved, so that the correlations in the $xy$ plane are fully isotropic -- a feature reproduced by the LSW approach, as we show in Appendix \ref{app-correl}. As illustrated in Fig. \ref{f.C_2D}, the spin correlations show an hybrid decay (exponential at short distance, followed by an algebraic decay at long distance):
			\bearr
				\langle S_i^y S_j^y \rangle &\sim & a e^{-r_{ij}/\xi} + 1 / r_{ij}^\alpha \\
				\langle S_i^z S_j^z \rangle - \langle S_i^z \rangle \langle S_j^z \rangle &\sim & \epsilon_i \epsilon_j \left ( a' e^{-r_{ij}/\xi} + 1 / r_{ij}^{2\alpha} \right ) 
			\eearr
			with $a$, $a'$ some ($\alpha$- and $\Delta$- dependent) constants.
	 The simple exponential decay expected in the finite-range regime is thus recovered only for $\alpha \to \infty$ in a rather nontrivial way, while for any $\alpha< \infty$ the spatial decay of interactions dictates the long-distance spatial decay of correlations. Similar findings were reported earlier on other gapped systems in Refs.~\cite{peteretal2012, vodolaetal2014}. The hybrid decay (exponential followed by a power law) is a common feature of gapped systems with long-range interactions -- it has been observed, for instance, in the dynamics following a quench \cite{gongetal2014}, and in the ground state of a spin-1 topological phase \cite{gongetal2016b}.

		\subsection{Fluctuations of the collective spin}
		\label{sec-fluctu}
		
%
		
		To complement the analysis of the large distance decay of the spin correlations, we offer in this Section a scaling analysis of the associated fluctuations of the collective spin. Considering the collective spin component $S^{\beta} = \sum_i S^{\beta}_i$ with $\beta = x, y, z$, one has 
		\be
		\langle (\delta S^{\beta})^2 \rangle = \sum_{i} \langle (\delta S^{\beta}_i)^2 \rangle
		+ \sum_{i\neq j} \langle \delta S_i^{\beta} \delta S_j^{\beta}  \rangle
		\label{e.deltaS}
		\ee  		
		(where we have introduced the symbol $\delta O = O - \langle O \rangle$), namely the fluctuations are composed of a local term and of a correlation term. If $\left[ S^{\beta},\cal H \right] \neq 0$ - namely for the non-conserved spin components $\beta = x,y$ of the collective spin, the ground state of the Hamiltonian has finite global fluctuations of the collective spin component in question, whose scaling can be readily estimated from the knowledge of the power-law decay of the correlation function $\langle \delta S_i^{\beta} \delta S_j^{\beta}  \rangle \sim r_{ij}^{-\eta_{\beta}}$. 
	Indeed, from Eq.~\eqref{e.deltaS} one readily deduces that:
	\be
	\langle (\delta S^{\beta})^2 \rangle \sim {\cal O}(L^d) + {\cal O}(L^d) \int_a^L dr ~ r^{d-1-\eta_\beta}~.
	\ee    
Clearly, if $d-1-\eta_\beta < -1$ ($\eta_\beta > d$) , then the scaling of fluctuations is ${\cal O}(L^d)$, while $d-1-\eta_\beta > -1$ ($\eta_\beta < d$) will provide a correction to the conventional volume law of fluctuation scaling, namely
\be
\langle (\delta S^{\beta})^2 \rangle \sim  L^{2d-\eta_\beta}~.
\ee 
In the N\'eel phase, $\eta_{x,y} = \alpha > d$, so that corrections to volume scaling of fluctuations do not appear.
		
\begin{figure}
		\includegraphics[width=\linewidth]{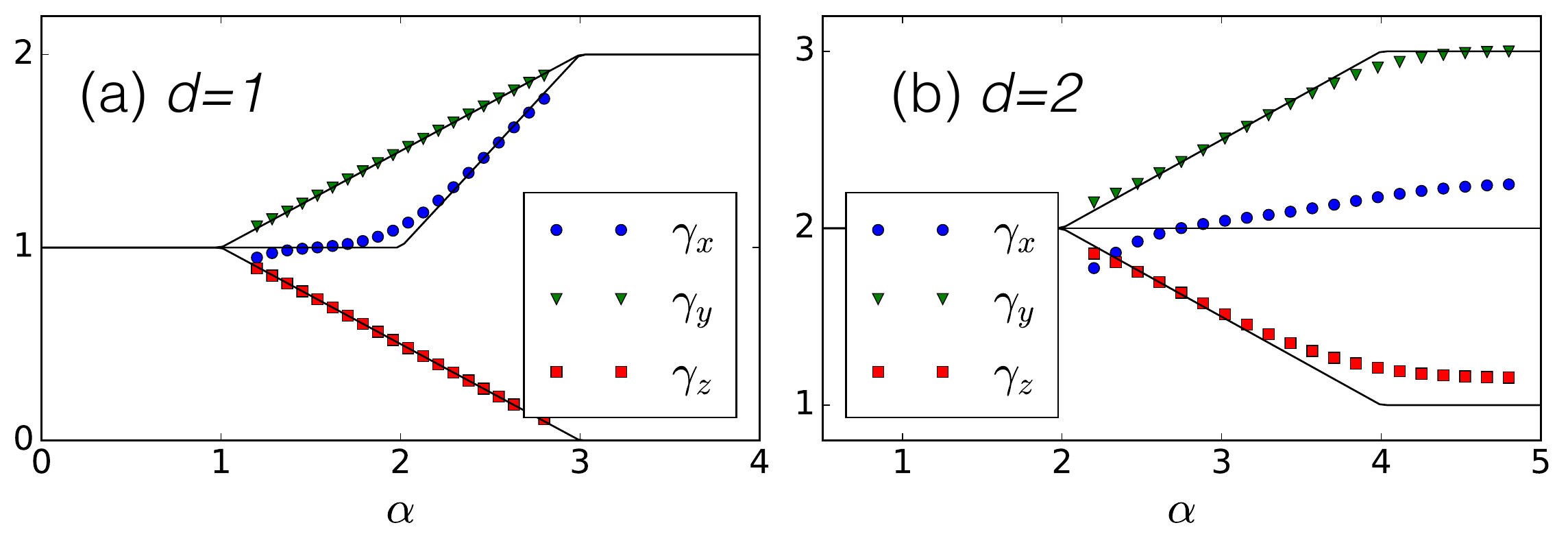}
		\caption{Scaling exponents of the collective spin fluctuations in a subsystem of linear size $L$: $\langle \delta^2 S_A^\beta \rangle \sim L^{\gamma_\beta}$ for $\beta = x,y,z$. Here we consider the XY phase for $d=1$ and 2. For $\alpha > d+2$, we find in fact a logarithmically violated area law for the $z$ component : $\langle \delta^2 S_A^z \rangle \sim L^{d-1} \log L$. Solid lines are the theoretical predictions of Table \ref{table_summary}, while the dots result from fits with $A$ half of the total system , which is a $L \times 2L$ torus in $d=2$ and a $2L$ circle in $d=1$. Fitting functions were chosen of the form $\langle \delta^2 S_A^{\beta} \rangle = a L^{\gamma} + b$. Sizes $L=10\dots 200$ in $d=2$ and $L=10^4 \dots 10^5$ in $d=1$ were used for the fits.}
		\label{f.exp_variance}
	\end{figure}

In the XY phase, on the other hand, given the results presented in Table \ref{table_correl_functions}, the whole medium-range regime is characterized by a violation of the volume law of fluctuations for the $S^y$ collective spin, 
$\langle (\delta S^y)^2 \rangle \sim L^{d+z}$, while the longitudinal fluctuations of the order parameter $\langle (\delta S^x)^2 \rangle \sim L^{\max(d,2z)}$ could violate a volume law only if $z \geq d/2$, implying $2d  < \alpha < d+2$ -- a condition which is only satisfied in $d=1$ for $2 < \alpha < 3$ (see Fig. \ref{f.exp_variance}(a)).

When considering instead the $S^z$ spin component, which is conserved by the XXZ Hamiltonian, one has that $\langle (\delta S^z)^2 \rangle = 0$, so that the only meaningful question to ask concerns the scaling of the fluctuations of the $S^z_A = \sum_{i \in A} S_i^z $ operator defined on subsystem $A$.  In particular the fluctuations of the local $S^z_A$ spin component can be expressed entirely in terms of correlations between $A$ and $B$, namely 
	\bearr
			\langle \delta^2 S^z_A \rangle &=& - \langle  \delta S^z_A \delta S^z_B \rangle \nonumber \\
			&=& - \sum_{i \in A} \sum_{j \in B} [\langle  S^z_i S^z_j \rangle - \langle  S^z_i\rangle \langle S^z_j \rangle ]~.
		\eearr	
The scaling of $\langle \delta^2 S^z_A \rangle$	is then fundamentally governed by the decay of correlations. Considering that the system is defined on a hyper-torus, and considering an equal $A-B$ bipartition of the system (namely $L_A, L_B \sim L$), one can show that
		\be
			\sum_{i\in A} \sum_{j \in B} \frac{1}{| r_i - r_j |^{\eta_z}} \sim 
			 \left\{
      \begin{aligned}
        L^{2d - \eta_z} & ~~\textnormal{if} & \eta_z < d+1 \\
        L^{d-1} \log L & ~~\textnormal{if} & \eta_z = d+1 \\
        L^{d-1} & ~~\textnormal{if} & \eta_z > d+1~. \\
      \end{aligned}
    \right.
    \label{eq_scaling_covar}
		\ee	
Details on the calculation are provided in Appendix \ref{app-scaling-covar}, showing that further logarithmic corrections are expected if $\eta_z=d$ or $\eta_z = d-1$.  Hence in the N\'eel phase, where $\eta_z  = 2\alpha > 2d \ge d+1$, an area-law scaling of $S^z$ fluctuations is verified over the whole medium- and short-range regimes in all dimensions $d\geq 1$.
	
In the XY phase, on the other hand, $\eta_z = d+z$, so that the whole of the short-range regime with $z=1$ exhibits a logarithmically violated area law for the $S^z$ fluctuations, as observed in the finite-range limit \cite{bipartite-fluct}; whereas the medium-range regime with $z<1$ exhibits a continuous violation of the area law, $\langle \delta^2 S^z_A \rangle \sim L^{d-z}$, up to a fully developed volume-law scaling in the long-range regime. 

Our findings for the scaling of fluctuations are summarized in Table~\ref{table_summary}. Fig. \ref{f.exp_variance} shows the predicted scaling exponents for the collective-spin fluctuations, in the XY phase compared with the numerical reconstruction of the exponents via direct calculations on systems with sizes  $L=10^4 \dots 10^5$ in $d=1$ and $L=10 \dots 200$ in $d=2$, showing that finite-size effects may be significant, yet they do not prevent from observing the strong $\alpha$ dependence of the scaling of fluctuations in the medium-range regime.


		\section{Scaling of the subsystem entanglement entropy}	
		\label{s_entanglement}	
			
			In this section we investigate the entanglement properties of the ground state, focusing on the scaling of the entanglement entropy of a subsystem. The latter is defined starting from the reduced density matrix (RDM) $\rho_A$ of subsystem $A$ as a partial trace of the ground-state projector over $B$ (namely the complement of $A$):			
			\be
			\rho_A = {\rm Tr}_B |\Psi_0\rangle \langle \Psi_0|~.
			\ee
	The entanglement entropy (EE) is then defined as the von Neumann entropy of the RDM,  $S_A = - {\rm Tr} \rho_A \log \rho_A$. 		
	It is useful to recall that any quantum state admits a so-called Schmidt decomposition \cite{Peres-book}  related to the $A$-$B$ bipartition of the system, in the form:
	  		\be 
	  			|\psi_{AB}\rangle = \sum_p \sqrt{\lambda_p}~ |\psi_A^{(p)}\rangle \otimes |\phi_B^{(p)}\rangle
	  		\ee
	  		where the states $|\psi_A^{(p)}\rangle$ (resp. $|\phi_B^{(p)}\rangle$) form an orthogonal basis of the Hilbert space $H_A$ (resp. $H_B$) of subsystem $A$ (resp. $B$). The EE  is then simply obtained as the Shannon entropy of the Schmidt coefficients $\lambda_p$ $S_A =-\sum_p \lambda_p \log \lambda_p$. 

	LSW theory allows for a very efficient calculation of the entanglement properties of the ground state, thanks to the Gaussian form of the RDM of any subsystem. We refer the reader to Sec.~\ref{sec-area_law_EE} and to Refs.~\cite{song,Luitzetal2015,FrerotR2015}, where the calculation of the EE for quadratic bosonic systems is detailed.	
			
	  		\subsection{Entanglement and fluctuations in the long-range regime}
			\label{s_EELR}
	 		
	  	When $\alpha<d $ the Schmidt decomposition of the exact ground state in the thermodynamic limit can be determined explicitly, allowing for an exact calculation of the EE \cite{Latorreetal2005}.  			
	  		We have argued in Section \ref{sec-infinite_range} that for $\alpha < d$ the ground state is, in the thermodynamic limit,  $| \Psi_{0} \rangle = |S = N/2,~ M = 0 \rangle$. Introducing $|S = N/2,~ M = -N/2 + p \rangle$ for the symmetric, equal-weight superposition of all states having $p$ spins up and $N - p$ spins down, the Schmidt decomposition of $| \Psi_{0} \rangle \propto (\sum_i \sigma_i^+)^{N/2}  |S_{A} = \frac{N_A}{2},~ M_A = -\frac{N_A}{2} \rangle \otimes |S_{B} = \frac{N_B}{2}, ~ M_B = -\frac{N_B}{2} \rangle$ for a bipartition into $N_A \le N / 2$ and $N_B = N - N_A$ spins is then simply 
	  		\bearr
	  			| \Psi_{0} \rangle = \sum_{p=0}^{N_A} & ~\sqrt{\lambda_p}~  |S_{A} = \frac{N_A}{2},~ M_A = -\frac{N_A}{2} + p \rangle \otimes \nonumber \\
	  			& |S_{B} = \frac{N_B}{2}, ~ M_B = -\frac{N_B}{2} + \frac{N}{2} - p \rangle
	  		\label{e.schmidt-MF}
	  		\eearr
	  		where $\lambda_p = \binom{N_A}{p} \binom{N_B}{N/2 - p} / \binom{N}{N/2}$ counts the number of ways (normalized to unity) in which to distribute $p$ up-spins up among $N_A$ spins, and $N/2 - p$ up-spins among $N_B$ spins --- the remaining $N/2$ spins being down-spins. The distribution of the subsystem magnetization $M_A = S_A^z$ is also given by $\lambda_p$, and it is centered around $S_A^z = 0$ with a width of order $\sqrt{N_A}$, hence a variance scaling as $N_A \sim L_A^d$, consistent with the scaling $L_A^{d-z}$ and $z=0$ (compare Table \ref{table_summary}).  
			
		The EE is simply the entropy of the $\lambda_p$ distribution, which scales as $(d/2) \log L_A$ ---indeed, the entropy is of order $\log \Omega$, with $\Omega \sim  \sqrt{N_A}$ the typical number of nonzero $\lambda_p$'s. In this long-range regime, the Schmidt basis   
$|\psi_A^{(i)}\rangle$  is contained in a small subspace of the local Hilbert space (namely that of symmetric superposition states with a fixed magnetization), having dimensions $O(N_A)$ to be contrasted with the local Hilbert space dimensions ($2^{N_A}$). The example at hand highlights the existence of a simple relation between the structure of ground-state entanglement and that of the fluctuations of the subsystem magnetization $S_A^z$, the latter being the only physical mechanism responsible for entanglement. Even though such a simple relationship cannot be found in the short- and medium-range regimes, one can always say that, if $S_z$ is globally conserved in the system, its fluctuations on subsystem $A$ are a sufficient condition for the existence of $AB$ entanglement \cite{FrerotR2015}.

		\subsection{Entanglement and fluctuations in the medium- and short-range regime}
		\label{sec-area_law_EE}
		
		 
		
\begin{figure}
			\includegraphics[width=1.\linewidth]{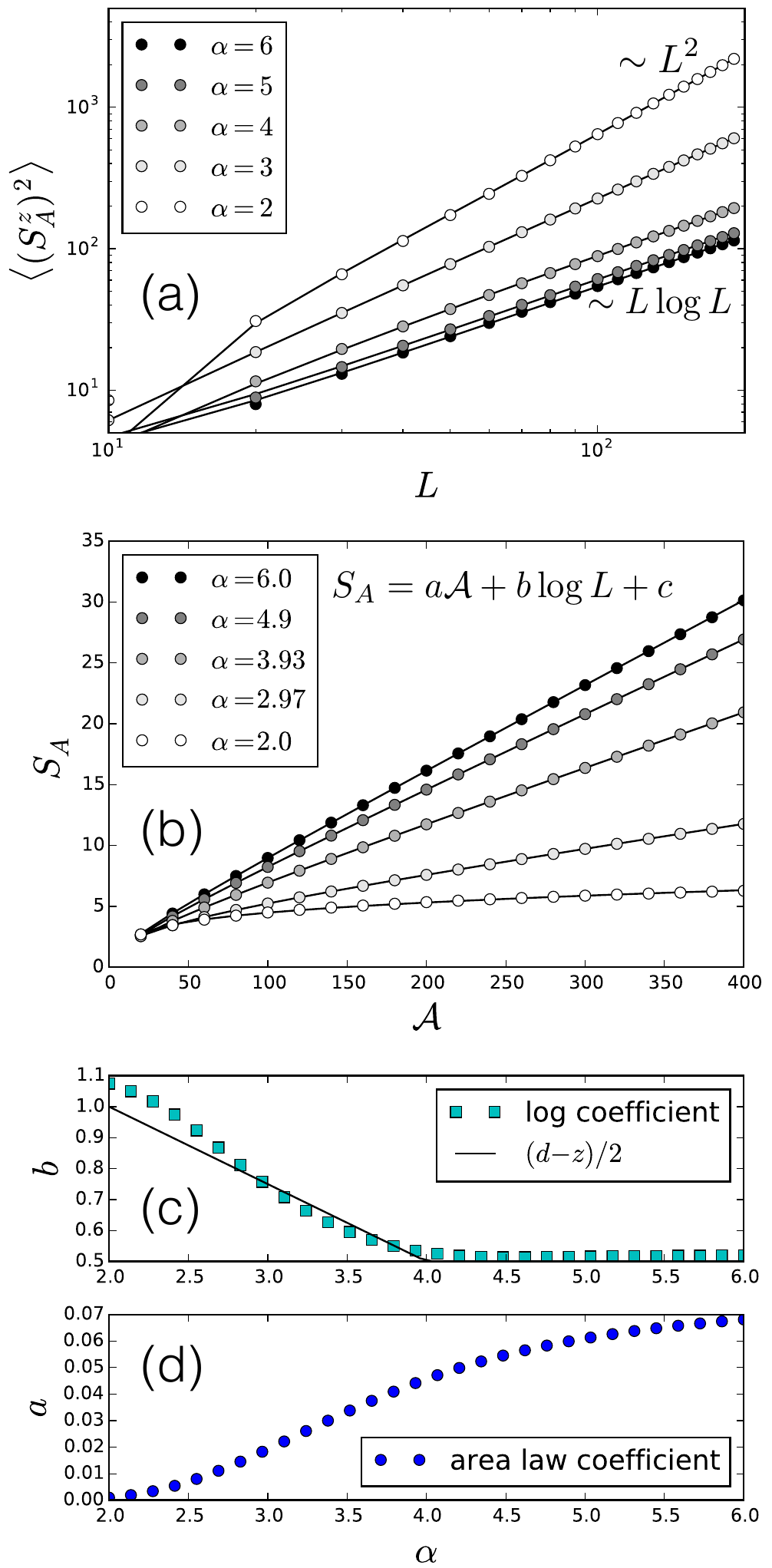}
			\caption{Area law of entanglement entropy (EE) and violation of the area law for $\langle (S_A^z)^2 \rangle$ in $d=2$ in the XY phase ($\Delta=0$). $A$ is half of a $2L \times L$ torus, with $L=10\dots 200$. (a) $\langle (S_A^z)^2 \rangle$ is plotted for various values $\alpha$ as a function of $L$. Solid lines are fits of the form $\langle (S_A^z)^2 \rangle = a L^{d-z} + b L^{d-1} + c$ when $\alpha < d+2$, and $\langle (S_A^z)^2 \rangle= a L^{d-1} \log L + b$ when $\alpha > d+2$. (b) EE $S_A$ for various values $\alpha$ as a function of the boundary area ${\cal A}$ ($=2 L$ in $d=2$). Solid lines are fits of the form $S_A = a {\cal A} + b \log L + c$. (c-d) $b$ and $a$ coefficients of the scaling of EE, plotted as a function of the decay exponent $\alpha$. The log-coefficient $b$ is compared to the prediction $b=N_G (d-z) / 2$, where $N_G=1$ is the number of Goldstone modes, $d=2$ the dimension of space, and $z$ the dynamical exponent.}
			\label{f.S_scaling_2d}
		\end{figure}

		In this section, we discuss the scaling of the EE of a subsystem in the medium- and short-range regime as obtained via LSW theory, contrasting it with the scaling of the fluctuations of the subsystem magnetization $\langle \delta^2 S_A^z \rangle$. Our findings concerning the scaling of EE are summarized in Table \ref{table_summary}. In the gapped N\'eel phase, both quantities are found to obey and area law \cite{RMP-area-laws}, namely to scale as the boundary of $A$, in both regimes. In the XY phase, on the other hand, there is a stark contrast between the scaling of entanglement and that of magnetization fluctuations. Indeed entanglement is always found to obey an area law in both medium- and short-range regimes, while fluctuations exhibit systematic violations of the area-law scaling, as discussed in Sec.~\ref{sec-fluctu}. In particular the contrast is very stark in the medium-range regime, in which the prefactor of the area-law scaling of entanglement vanishes progressively as $\alpha$ is reduced towards $d$ (see Fig.~\ref{f.S_scaling_2d}(b)), while the scaling of $S_A^z$ fluctuations moves from an area-law one at $\alpha = d+2$ to a volume-law one at $\alpha = d$ [see Fig.~\ref{f.S_scaling_2d}(a)].
	
		In Fig. \ref{f.S_scaling_2d}(b) we fitted the EE as 
	\be
		S_A = a L_A^{d-1} + b \log L_A + c
	\ee
	with an additional subdominant logarithmic contribution $b \log L$, and a constant term $c$. Fig. \ref{f.S_scaling_2d}(d) shows that the area-law coefficient decreases monotonically to zero when $\alpha$ decreases towards $d$. The $b$ coefficient, shown in Fig. \ref{f.S_scaling_2d}(c), can be attributed to the Goldstone modes associated with the broken rotational symmetry \cite{metlitskiG2011}, and it will be discussed in Section \ref{sec-toy_model}, while the special case of $d=1$ (possessing a logarithmic scaling even in the absence of long-range order) will be analyzed in Sec.~\ref{sec-TOS_1d}. 

	\begin{figure}
			\includegraphics[width=1.\linewidth]{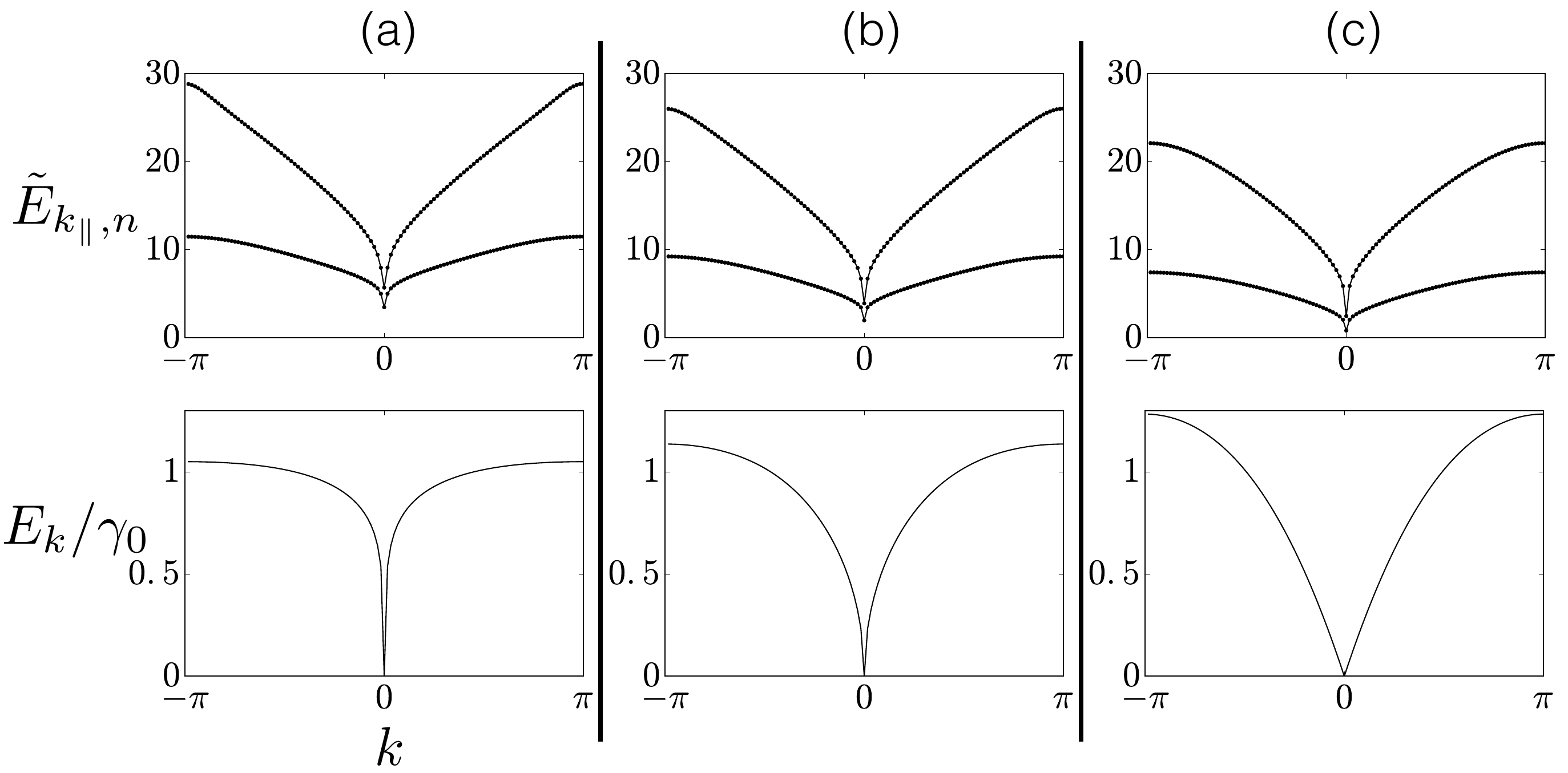}
			\caption{Entanglement spectrum (top line) versus physical spectrum (bottom line) in $d=2$ at $\Delta=0$ for $\alpha=5$ (a), $\alpha=3$ (b) and $\alpha=2.2$ (c) (XY phase). Entanglement spectrum is plotted as a function of the momentum $k_\parallel$ parallel to the boundary cut between $A$ and $B$ (only the two lowest branches are shown), and physical spectrum along the diagonal $(k,k)$ of the Brillouin zone. The apparent gap at $k_\parallel=0$ in the entanglement spectrum scales to zero upon increasing the system size \cite{FrerotR2015}.}
			\label{f.ES_vs_PS_Jz_0}
		\end{figure}

The strong decrease of the EE upon decreasing $\alpha$ can be understood from the structure of the single-particle entanglement spectrum, and from its comparison to the (bulk) physical spectrum. The single-particle entanglement spectrum is defined starting from the Gaussian structure of the RDM $\rho_A$ describing the $A$ subsystem, which in turn is a consequence of the harmonic approximation made within LSW theory \cite{FrerotR2015}:
\be
\rho_A =: \exp(-{\cal H}_A) =  \exp \left[ -(\bm b^{\dagger}, \bm b)^{T} ~h_A~ (\bm b, \bm b^{\dagger}) \right] 
\ee 
where ${\cal H}_A$ is the so-called entanglement Hamiltonian, ${\bm b} = (b_1, b_2, ..., b_{N_A})$ is the vector of Bose operators on the $A$ subsystem, and $h_A$ is the single-particle entanglement Hamiltonian. The diagonalization of $h_A$ \cite{FrerotR2015} brings the quadratic entanglement Hamiltonian to the form ${\cal H}_A = \sum_{\bm k_{||},n} \tilde{E}_{\bm k_{||},n} \beta_{\bm k_{||},n}^{\dagger} \beta_{\bm k_{||},n}$, where $\bm k_{||}$ is the momentum along the $A$-$B$ cut -- representing a good quantum number for the half-torus geometry of subsystem $A$ that we adopt here -- and $n$ is a further mode index related to the dynamics generated by ${\cal H}_A$ transversely to the cut. 

	\begin{figure}
			\includegraphics[width=1.\linewidth]{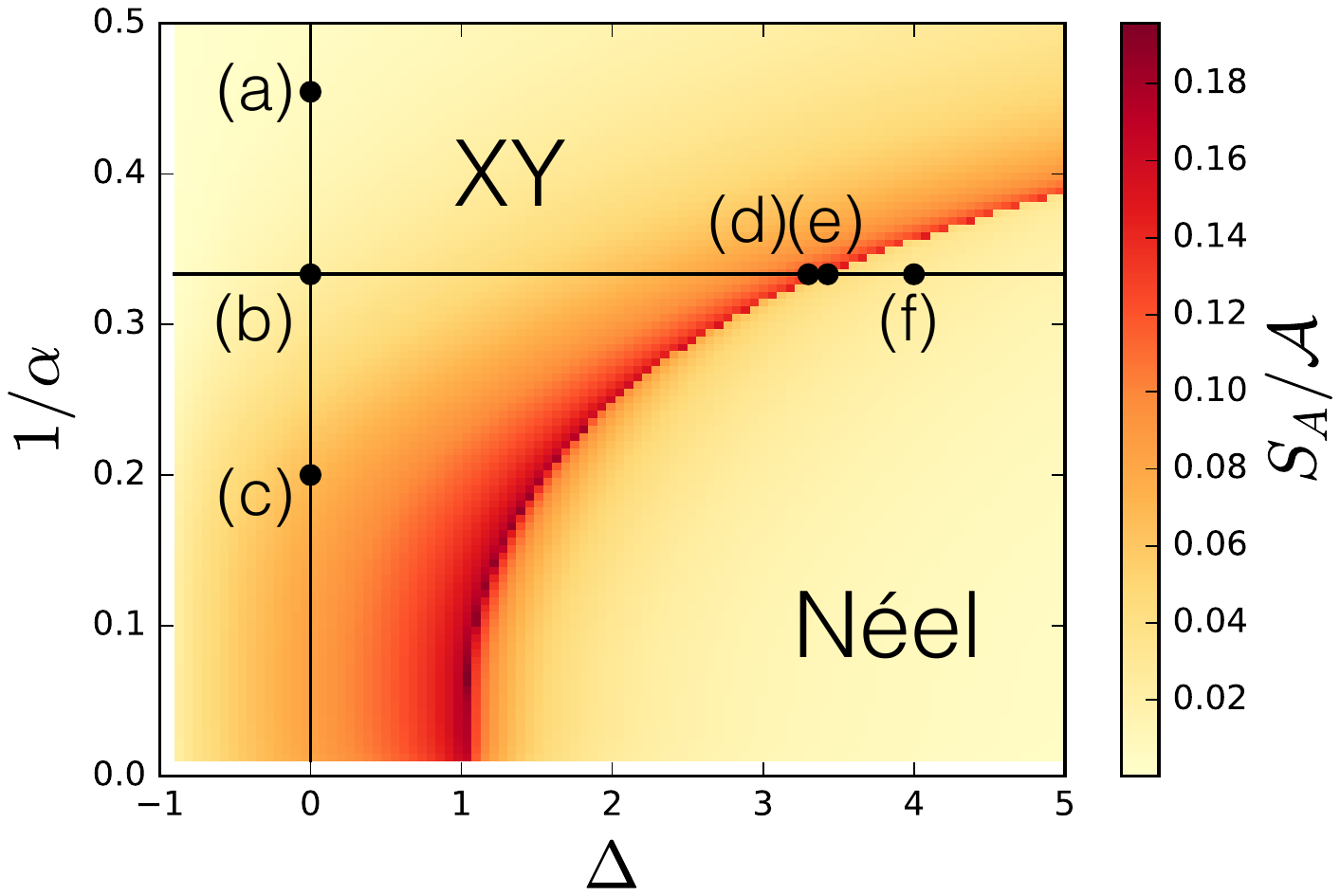}
			\caption{Entanglement entropy $S_A$ above the ground-state phase diagram of the 2D-XXZ Hamiltonian. $A$ is half of a $60 \times 30$ torus. Dots indicate the points where the entanglement spectrum and physical spectrum are compared on Fig. \ref{f.ES_vs_PS_Jz_0} and Fig. \ref{f.ES_vs_PS_alpha_3}.}
			\label{f.S_phase_diagram_2D}
		\end{figure}

\begin{figure}
			\includegraphics[width=1.\linewidth]{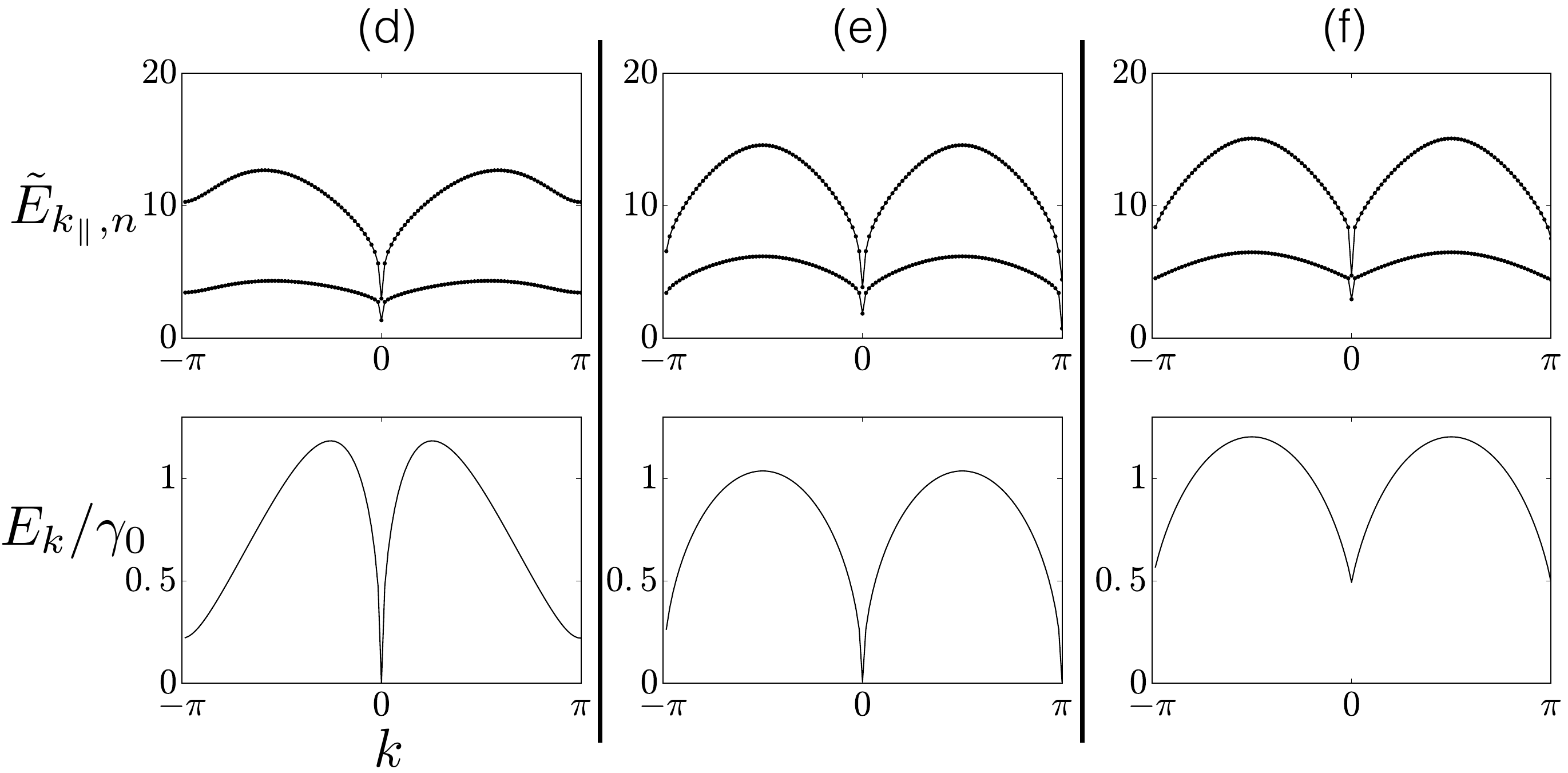}
			\caption{Entanglement spectrum (top line) versus physical spectrum (bottom line) in $d=2$ at $\alpha=3$ for $\Delta=3.3$ (d), $\Delta=3.429$ (e) and $\Delta=4$ (f). Entanglement spectrum is plotted as a function of the momentum $k_\parallel$ parallel to the boundary cut between $A$ and $B$ (only the two lowest branches are shown), and physical spectrum along the diagonal $(k,k)$ of the Brillouin zone. The apparent gap at $k_\parallel=0$ in the entanglement spectrum in (d1) and (e1) scales to zero upon increasing the system size \cite{FrerotR2015}.}
			\label{f.ES_vs_PS_alpha_3}
		\end{figure}

The Gaussian RDM $\rho_A$ describes a thermal gas of quasiparticles with dispersion relation given by $\tilde{E}_{\bm k_{||},n}$ at unit temperature - whose thermal entropy is identical to the EE. 
 Fig. \ref{f.ES_vs_PS_Jz_0} shows that the low-lying modes of the single-particle entanglement spectrum $\tilde{E}_{\bm k_{||},n}$ become increasingly stiff as $\alpha$ decreases -- in striking analogy with the physical spectrum of spin-wave excitations, whose low-energy density of states behaves as $\rho(\omega) \sim \omega^{d/z - 1}$, becoming smaller when $\alpha$ (and $z=(\alpha - d)/2$) decrease. A progressive reduction in the low-energy density of states of the single-particle entanglement spectrum $\tilde{E}_{\bm k_{||},n}$ obviously implies a reduction of entropy at fixed temperature for $\rho_A$, namely of EE.  Conversely, the EE is maximal along the phase-transition line between the XY and N\'eel phase, as shown on Fig. \ref{f.S_phase_diagram_2D}, and this increase can be again  associated with the $\tilde{E}_{\bm k_{||},n}$ spectrum, which, at the transition, acquires a second zero mode at  $\bm k_{||}=(\pi,\pi,...)$, besides the one already present at $\bm k_{||}=0$, as shown in Fig.~\ref{f.ES_vs_PS_alpha_3}. The appearance of a second soft mode obviously boosts the low-energy density of states, leading to an increase of the EE \cite{FrerotR2016}. Correspondingly, the spin-wave spectrum (also shown in Fig.~\ref{f.ES_vs_PS_alpha_3}) shows the same appearance of a second soft mode (at $\bm K = (\pi,\pi,...)$), which signals the instability of the XY ground state to the appearance of long-range staggered spin order as in the N\'eel phase. The corresponding increase in the low-energy density of states of the spin-wave excitations implies in turn a stronger quantum correction of the classical ordered moment: hence the simultaneous softening of the spin-wave dispersion relation and of the single-particle entanglement spectrum at the transition is responsible for the striking similarity between the strong enhancement of entanglement  (Fig. \ref{f.S_phase_diagram_2D}) and of quantum fluctuations of the order parameter (Fig. \ref{f.phase_diagram_2d}) around the XY-N\'eel transition. 
  

		
\begin{figure}
			\includegraphics[width=1.\linewidth]{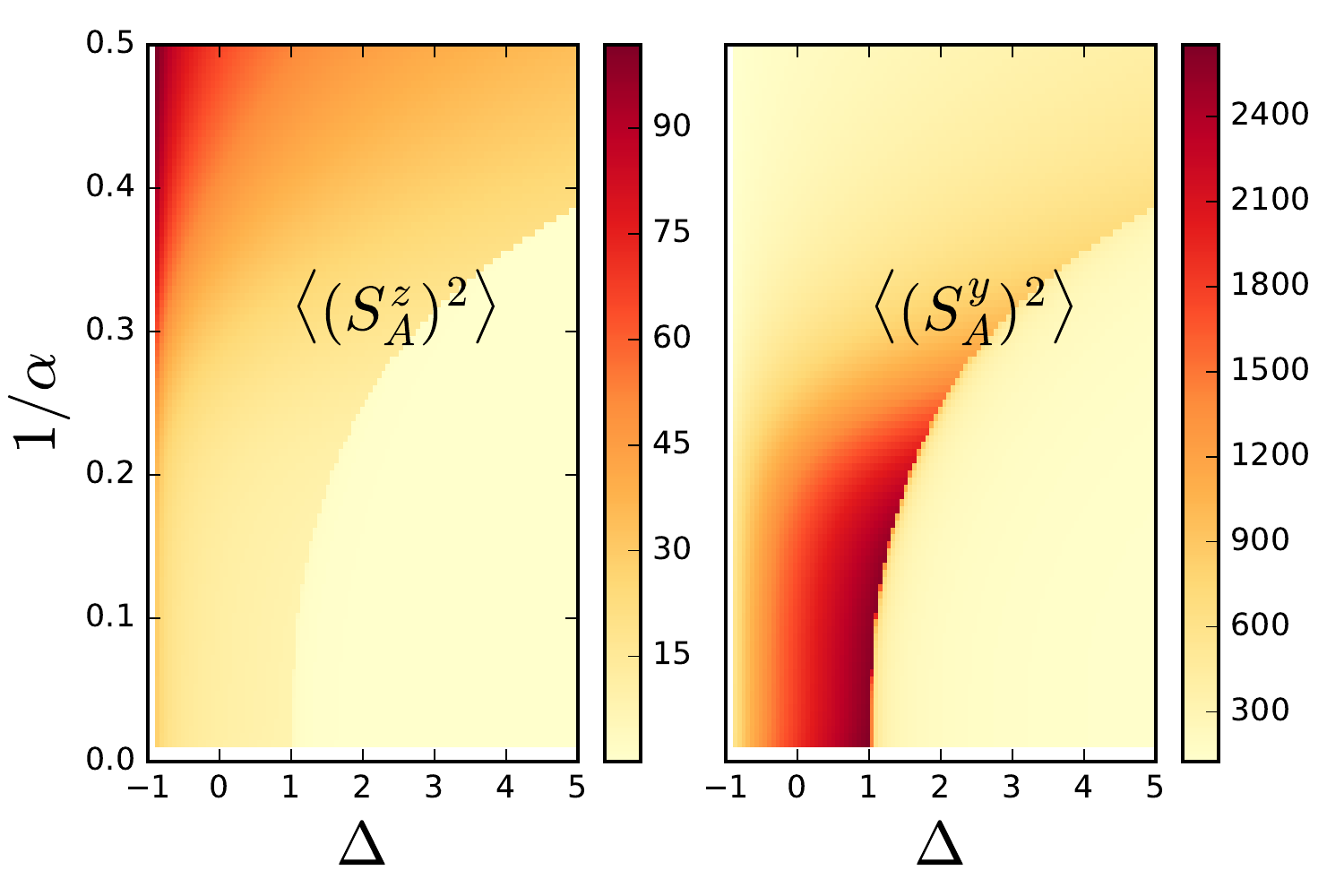}
			\caption{Variance of the $z$ (left) and $y$ (right) component of the total spin of a subsystem $A$ throughout the ground state phase diagram of the 2D-XXZ Hamiltonian. Geometry as in Fig. \ref{f.S_phase_diagram_2D}.}
			\label{f.varSz_varSy_phase_diagram_2D}
		\end{figure}

Finally, Fig. \ref{f.varSz_varSy_phase_diagram_2D} shows the variance of $S_A^z$ and of $S_A^y$ across the phase diagram (for a fixed finite size). The evolution of spin fluctuations along the $y$ axis strongly resembles that of the entanglement shown in Fig.~\ref{f.S_phase_diagram_2D}, although it more clearly reveals the first-order nature of the transition with a jump in the fluctuations properties. The jump is related to a sudden change in the scaling properties of the fluctuations, as detailed in Table \ref{table_summary}. A jump is also to be observed in the EE, albeit much weaker as the dominant scaling behavior of entanglement is the same area law on both sides of the transition.

		
		\subsection{Medium- and short-range regime in the XY phase: sub-leading contribution to the entanglement entropy}
		\label{sec-TOS}
		We now take a closer look at the details of the entanglement scaling in the XY phase, focusing on the observed logarithmic corrections to the area law scaling of EE - as reported in the previous section.  We first recall in Section \ref{sec-toy_model} the origin of this logarithmic contribution in the presence of a broken symmetry, as stemming from the interplay between the tower-of-state entanglement spectrum of a subsystem and the low-lying Goldstone modes coupling two subsystems.  In so doing we shall rephrase arguments which have been put forward in earlier works \cite{metlitskiG2011, kulchytskyyetal2015}, but we will also generalize them to the case of long-range interactions, which add the new ingredient of a continuously varying dynamical exponent $z\leq 1$. Furthermore we shall specialize in Section \ref{sec-TOS_1d} our discussion to the case of $d=1$, where the analysis is somewhat more subtle. There the logarithmic correction to the area law becomes the dominant term in the medium-range regime, and interestingly a similar logarithmic scaling persists also in the short-range regime, featuring a Luttinger-liquid (LL) phase with conformal invariance: supplementing our LSW calculations with numerical DMRG data will allow us to study the evolution of entanglement across the XY-LL transition. 
		
		\subsubsection{XY phase: logarithmic term from the tower-of-state spectrum}
		\label{sec-toy_model}
	
 A universal additive logarithmic contribution to the area-law scaling of EE has been predicted to appear in systems breaking a continuous symmetry in their ground state in Ref.~\cite{metlitskiG2011}, and extensively verified numerically in Refs.\cite{Luitzetal2015,kulchytskyyetal2015,FrerotR2015}. Such a contribution can be traced back to the existence of a low-lying tower of states (ToS) in the entanglement spectrum of a subsystem $A$, akin to the low-lying spectrum of a finite-size system exhibiting spontaneous symmetry breaking in the thermodynamic limit \cite{Anderson-book}. If isolated from its complement $B$, subsystem $A$ would indeed possess such a ToS in the low-lying excitation spectrum, described by the effective Hamiltonian $H^{\rm (ToS)} = {\cal L}^2 / 2I$. Here $\cal L$ stands for the generator of the symmetry spontaneously broken in the thermodynamic limit -- in the case of XY symmetry at hand ${\cal L} = S_A^z$, the generator of rotations in the $xy$ plane. The ToS Hamiltonian describes therefore the angular momentum of a rigid rotor living on a $N$-dimensional sphere, where $N$ is the number of components of the order parameter ($N=2$ in our case), and possessing a moment of inertia $I \sim L_A^d$, scaling as the volume of the subsystem. The exact same observations apply to subsystem $B$. 
 
  The coupling between subsystems $A$ and $B$ creates entanglement between the $A$ and $B$ rigid rotors in the joint ground state. As a consequence of the $A$-$B$ coupling the system possesses $N_G = N-1$ Goldstone modes spreading coherently across the two subsystems. In particular the characteristic energy scale for the coupling of the $A$ and $B$ rigid rotors is set by the smallest-wavevector Goldstone mode(s) with $k\sim 1/L$, to which we attribute an energy $\Delta_G = \Delta_G(L)$. As discussed in Ref.~\cite{metlitskiG2011, kulchytskyyetal2015}, the Hamiltionan of the coupled $A$ and $B$ rotors can then be approximated as that of a harmonic oscillator of frequency $\Delta_G/\hbar$ \cite{metlitskiG2011, kulchytskyyetal2015}. Tracing out subsystem $B$ leads to a density-matrix description of subsystem $A$, in which ToS modes are populated up to an energy of the order of $\Delta_G$, namely up to an angular momentum ${\cal L}_{\max} \sim (I \Delta_G)^{1/2}$; hence, knowing that ToS levels ${\cal L}$ have a degeneracy of order ${\cal L}^{N-2}$, the EE can be estimated by simple state counting as $S_A \sim \log \Omega_A$, where 
  \be
  \Omega_A \sim \int_0^{{\cal L}_{\rm max}} {\cal L}^{N-2}d{\cal L}  = (I \Delta_G)^{(N-1)/2}~. 
  \ee
   As a consequence the EE receives a contribution from the ToS spectrum of the kind
 \begin{equation}
  S_{\rm ToS} = \frac{N-1}{2} \log (L_A^d \Delta_G) + {\rm const.}
  \end{equation}
 As discussed at length in Sec.~\ref{s.spectrum} in the long-range XXZ model of interest here $\Delta_G \sim L^{-z}$, leading then to the result
  \begin{equation}
  S_{\rm ToS} = \frac{N_G(d-z)}{2} \log L_A + {\rm const.}
  \end{equation}
 The above result generalizes to the case $z\neq 1$ the universal logarithmic term in the EE of a continuous-symmetry-breaking phase first obtained in Ref.~\cite{metlitskiG2011}.  
 			
			\begin{figure*}
			\includegraphics[width=1.\linewidth]{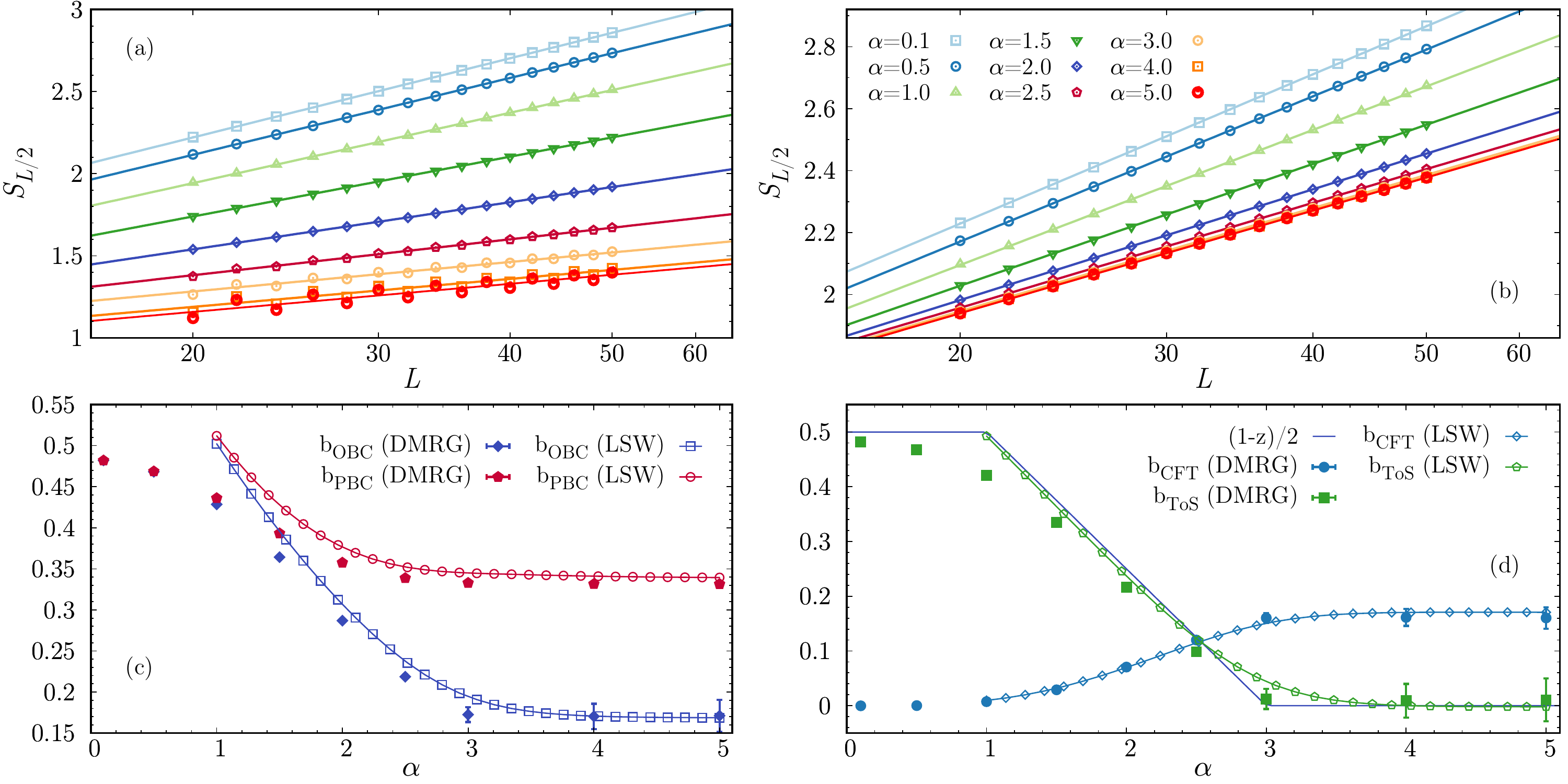}
			\caption{Scaling of half-chain entanglement entropy determined via DMRG simulations for different system sizes $L=20, 22 \dots 50$ and $\Delta=0$. Simulations have been performed for both OBC (a) and PBC (b). (c) The coefficients $b_{\rm OBC}$ and $b_{\rm PBC}$ result from fitting the entanglement entropy as $S_{L/2}=b \log L + c$. For the LSW data, we have used the same system sizes as in the DMRG calculations to fit the coefficients (having added a small transverse field term $- S_{\rm tot}^x / L^2$). The LSW coefficients show very little finite-size effects, and they are not found to change significantly when extending the size of the system up to $L=10^5$. (d) Extraction of $b_{\rm CFT} = b_{\rm PBC} - b_{\rm OBC}$ and $b_{\rm ToS} = 2b_{\rm OBC} - b_{\rm PBC}$ and comparison with the theoretical prediction $b_{\rm ToS} = (1-z)/2$ (black solid line, see Text).}
			\label{f.SscalingDMRG}
		\end{figure*}
		
  Within LSW theory there is no guarantee that this additive logarithmic contribution can be accurately captured, especially in view of the fact that the LSW approach cannot describe the ToS sector of the spectrum, as it assumes an explicitly broken symmetry even on finite-size systems. Nevertheless, in the case of linearly dispersing ($z=1$) Goldstone modes it has been shown \cite{song,FrerotR2015} that a careful treatment of the $k=0$ sector of the LSW Hamiltonian enables one to actually capture the universal logarithmic term within LSW theory. This is possible by gapping out the $k=0$ LSW mode with the addition of a small term $- B S_{\rm tot}^x$ in the Hamiltonian which stabilizes the ferromagnetic order and introduces a gap of order $\sqrt{B}$ at $k=0$. The choice $B = 1/L^{2d}$ allows therefore to  mimic the $1/L^d$ scaling of the ToS spectrum in the thermodynamic limit. As shown on Fig.~\ref{f.S_scaling_2d}(b,c), this procedure leads to the appearance of a logarithmic term $b \log L_A$ in the EE extracted from the LSW data, whose $b$ prefactor reproduces rather closely the predicted universal form $b = (d-z) / 2$. The deviations can be in part attributed to finite-size effects, which, for a given maximum size $L$ used in the fits of the entanglement scaling, become more significant the lower $\alpha$, justifying the modest agreement, observed at low $\alpha$ in Fig.~\ref{f.S_scaling_2d}(c), between finite-size LSW calculations and the universal prediction.

			\subsubsection{From Luttinger liquid behavior to the XY phase in $d=1$}
			\label{sec-TOS_1d}
			
			As already commented above, the case $d=1$ is rather special in that the system possesses two ground-state phases in the XY regime upon varying $\alpha$: an XY phase with long-range order for $\alpha < 3$, and a Luttinger-liquid (LL) phase with algebraic correlations for $\alpha > 3$. In the latter regime the 
low-energy physics of the system is captured by conformal field theory (CFT) \cite{calabreseC2004}, predicting a universal logarithmic violation of the entanglement area law in the form $S_A = (n c/6) \log L_A$, where $c=1$ is the central charge of the CFT, and $n$ ($=1,2$) is the number of common boundaries between subsystem $A$ and its complement $B$. On the other hand, for $\alpha < 3$ conformal invariance breaks down (as $z<1$) and the system develops long-range order. The EE still possesses a universal logarithmic term, as discussed in the previous section, that is independent of the geometry of the $A$-$B$ boundary. It becomes therefore extremely interesting to explore the evolution of the EE across the XY-LL transition to track how the CFT prediction and the ToS contribution evolve into each other. 

LSW theory is obviously inadequate to the latter scope, as it breaks down for $\alpha \geq 3$ (and it becomes increasingly inaccurate as $\alpha \to 3^-$). Therefore we complement our approach with ground-state DMRG calculations \cite{whitedmrg, white93}, a method that gives easily access to the EE of a partition of the system. We perform the numerical simulations on lattices up to $L = 50$ sites implementing both open (OBC) and periodic boundary conditions (PBC) and setting $\Delta=0$. Since the Hamiltonian Eq.~\eqref{e.H_XXZ} commutes with total magnetization we restrict our study to zero magnetization sector. In order to keep a truncation error smaller than $10^{-7}$, throughout the simulations we used up to $M = 300$ states. We chose a hopping amplitude of the form $1 / |r_i - r_j|^\alpha$ for both OBC and PBC, both in the DMRG and the LSW calculations.\\
The scaling of half-chain EE is shown in Fig.~\ref{f.SscalingDMRG} for OPC and PBC in panels (a) and (b) respectively for different values of $\alpha$ ranging from $0.1$ to $5$. A general form for the dominant term of the EE scaling reads 
\be 
				S_A = (n ~b_{\rm CFT} + b_{\rm ToS})\log L_A + ...
			\ee
 where one expects $b_{\rm CFT} = c/6$ and $b_{\rm ToS} = 0$ for $\alpha > 3$,  $b_{\rm CFT} \neq 0$ and $b_{\rm ToS} = (d-z)/2=(3-\alpha)/4$ for $1 < \alpha < 3$, and $b_{\rm CFT} = 0$, $b_{\rm ToS} = 1/2$ for $\alpha < 1$. The meaning of $b_{\rm CFT}$ in the $1 < \alpha < 3$ (which lacks conformal invariance) is simply that of the prefactor to a logarithmic term which depends on the number of boundaries $n$.
The ability of both LSW theory \footnote{The introduction of OBC in LSW theory requires to diagonalize the quadratic Hamiltonian in real space - a well-established procedure which we implement following Ref.~\cite{blaizot-ripka}.} and DMRG to simulate both OBC ($n=1$) and PBC ($n=2$) allows to systematically extract finite-size estimates of the $b$ coefficients as  
		\be 
			\left\{
      \begin{aligned}
       b_{\rm CFT} &=& b_{\rm PBC} - b_{\rm OBC} \\
       b_{\rm ToS} &=& 2 ~b_{\rm OBC} - b_{\rm PBC} \\
      \end{aligned}
    \right.
		\ee		
The coefficients so extracted are shown in Fig. \ref{f.SscalingDMRG}: there we find a relatively good agreement between the theoretical expectations - valid in the infinite-size limit - and the finite-size numerical data, coming both from the quadratic LSW approximation, and from exact DMRG calculations. Finite-size effects appear to be very significant (and especially so at small $\alpha$). Yet the application of a transverse field $B \sim 1/L^2$ allows to use LSW theory on finite sizes even in the regime $\alpha >3$, and it is remarkable to observe that even in this regime LSW theory on finite system sizes remains predictive when compared with DMRG.


	\section{Conclusions and perspectives}
	\label{s_conclusions}
	Inspired by the possibilities of quantum simulation in ultracold atoms, we studied the low energy properties of the $d$-dimensional XXZ Hamiltonian with ferromagnetic couplings in the $xy$ plane and both ferro- and antiferromagnetic along the $z$ axis, both decaying as a power law $1/r^\alpha$ with the distance. Linear spin-wave theory proved to be a reliable tool to determine the ground state phase diagram, the low energy excitation spectrum as well as the correlation and entanglement properties in the various phases exhibited by the system: a gapped antiferromagnetic N\'eel ordered phase, a gapless ferromagnetic XY phase, and a gapped ferromagnetic Ising phase. 
	
	In the gapped N\'eel phase (present only for $\alpha > d$), we identified two fundamentally different regimes of interactions: a short-range regime ($\alpha > d+1$), and a medium-range regime ($d < \alpha < d+1$). Both regimes are similar in most aspects to the $\alpha \to \infty$ (nearest-neighbor) limit, with the main qualitative difference that the spin-spin correlations exhibits a short-distance exponential decay controlled by the gap $\Delta_g$ followed by a long-distance power-law decay of the correlations in the ground state, controlled directedly by the exponent $\alpha$. Furthermore, in the medium-range regime ($d < \alpha < d+1$) the short-range regime dispersion relation $E_k \sim \Delta_g + ck^2$ acquires a cusp $E_k \sim \Delta_g + ck^{\alpha-d}$ at small $k$, while the short-range dispersion relation is recovered at $\alpha > d+2$.
	
	In the XY phase, our calculations identify three fundamentally distinct regimes of interactions: a short-range regime ($\alpha > d+2$) characterized by a dynamical exponent $z=1$ akin to the finite-range limit ($\alpha=\infty$), a medium-range regime ($d < \alpha < d+2$) with a $\alpha$-dependent dynamical exponent $z=(\alpha -d)/2$, and a long-range regime ($\alpha < d$) exhibiting the same properties as the infinite-range regime ($\alpha = 0$) in the thermodynamic limit, namely correlations uniquely stemming from finite-size effects. In all regimes the dynamical exponent $z$ is found to control directly the long-distance decay of the spin correlations and the scaling of fluctuations of the collective spin in the ground state. In particular, the medium-range regime of the XY phase exhibits a continuously varying palette of scalings for the collective-spin fluctuations, all of them violating the scalings exhibited in the conventional short-range regime. In particular the local fluctuations of the conserved collective spin ($S^z$) on a subsystem exhibit algebraic corrections to the area law, as a result of the coherent exchange of magnetization quanta between two subsystems mediated by the non-local couplings. 
	
	The scaling of fluctuations is in sharp contrast with the behavior of entanglement entropy, which in the XY phase is found to always satisfy an area law (plus additive logarithmic corrections), and with an area-law prefactor which generically decreases when the range of interactions increases. This behavior can be traced back to the density of states in the quasi-particle entanglement spectrum (as well as in the physical spectrum), which is lowered upon decreasing $\alpha$ as the dispersion relation of quasi-particle excitations stiffens. Finally, an additive logarithmic correction to the area-law scaling is found in the XY phase, associated with the spontaneous breaking of the continuous U(1) symmetry. The prefactor to the logarithmic term is universal, uniquely depending on the $z$ exponent and the number of components of the order-parameter; our prediction generalizes in a non-trivial manner that of Ref.~\cite{metlitskiG2011} for short-range interacting systems, and it can be quantitatively tested against accurate density-matrix renormalization group results in $d=1$. 
	
 Beside their intrinsic theoretical interest, our results have a direct relevance for ongoing experiments on ultracold atoms in optical lattices. Indeed the Hamiltonian \eqref{e.H_XXZ} with $\alpha = 3$ in $d=2$ is realized by the Mott insulating phase of magnetic atoms \cite{dePazetal2013} when imposing the conservation of the magnetization along the quantization axis. Furthermore its $d=1$ implementation can be envisioned in trapped ions \cite{PorrasC2004, Monroeetal2014}, which also enable to vary continuously the $\alpha$ exponent of the power-law decay of interactions. In particular the existence of a medium-range regime exhibiting an $\alpha$-dependent, ``curved" dispersion relation of elementary excitations, as well as an $\alpha$-dependent scaling of fluctuations, represent a prediction which lend itself rather naturally to an experimental test. In the case $\alpha=3$, generalizing the model to antiferromagnetic interactions in the $xy$ plane (while taking $\Delta = 0$) would mimic the physics of lattice-trapped Rydberg atoms with resonant interactions \cite{barredoetal2015, Browaeysetal2016}. The XX model with nearest-neighbor ($\alpha=\infty$) interactions on a bipartite lattice has the same physics irrespective of the sign of the interactions. Therefore, in the case $\alpha>d$ - for which nearest-neighbor interactions dominate the energetics of the system - we may expect that much of the physics observed in the ferromagnetic case carries over to the antiferromagnetic one, as long as one restricts to bipartite lattices. The exploration of antiferromagnetic $xy$ interactions on bipartite lattices, as well as on non-bipartite ones, represents an exciting extension of our present work. 
 
 Our work shows that long-range interactions offer a very rich landscape in terms of scaling properties of ground-state quantum fluctuations and subsystem entanglement entropy. Establishing a quantitative link between the two is necessary to connect entanglement to directly measurable properties, yet it appears rather challenging. A strategy we shall pursue in the future is based on the concept of ``local entanglement thermodynamics"  \cite{swingleM2016}, postulating an explicit Ansatz for the entanglement Hamiltonian in the form of the original microscopic Hamiltonian of the system, yet with spatially modulated coupling constants.  Finally, the existence of sharply distinct regimes for the dispersion relation of elementary excitations in the model of interest suggests that the non-equilibrium unitary dynamics following a quench will also be extremely rich, revealing an unconventional spreading of correlations which directly reflects the existence of a continuously varying dynamical exponent $z$ \cite{Frerotetal2017}.

	\section{Acknowledgements}
	
We thank D. Vodola for fruitful discussions. PN acknowledges the Institut Universitaire de France and the INFN grant QUANTUM for financial support. This work was supported by ANR ("ArtiQ" project).

\appendix 

{
\section{Fourier transform of the interaction potential}
\label{app-gamma_k}
	In this Section, we prove the scaling behavior of Eq. \eqref{e.scaling-Jk}. To this goal, we analyze the small-$k$ behavior of the integral : 
	\be
	 	\gamma_{\bm k} - \gamma_{\bm k}^{(\rm n.n)} =  a^{\alpha - d} \int_{r > a} d^d {\bm r} ~ \frac{e^{i {\bm k}\cdot {\bm r} }}{r^\alpha}
	 \ee
	 for $\alpha > d$. More precisely, we will analyse $\gamma_0 - \gamma_{\bm k}$ at small $k$. Since $\gamma_0^{(\rm n.n)} -  \gamma_{\bm k}^{(\rm n.n)} = 2 \sum_{i=1}^d [1 - \cos(k_ia)] \sim (ka)^2$ on the hypercubic lattice we consider, $\gamma_0 - \gamma_{\bm k}$ will always scale at least as fast as $(ka)^2$ at small $k$.  Note that, although we can obtain the correct scaling for $\gamma_0 - \gamma_{\bm k}$, we are not able to predict the correct prefactors, as they depend on the details of the lattice which are not captured by our continuous approximation. Finally, we note that if $\alpha < d$:
	 \be 
	 	\gamma_k \sim (ak)^{\alpha-d} ~~~~(\alpha < d)
	 \ee
	 
	\subsection{$d=1$}			
	 In $d=1$, one has to evaluate : 
	 \be 
	 	a^{\alpha - 1} \int_a^\infty dr ~ \frac{2 \cos(kr)}{r^\alpha} = 2(ak)^{\alpha-1} \int_{ak}^\infty dx~ \frac{\cos x}{x^\alpha} ~.
	 	\ee
		As $\alpha > 1$, the integral diverges at $k \to 0$. One thus integrates by parts to obtain : 
		\be
			\int_{ak}^\infty dx ~ x^{-\alpha} \cos x = \left[\frac{x^{1 - \alpha}}{1 - \alpha} \cos x \right]_{ak}^\infty + \int_{ak}^\infty dx ~ \frac{x^{1 - \alpha}}{1 - \alpha} \sin x ~.
		\label{e.fourier_pot_1d}
		\ee
		Now, since $1 - \alpha < 0$, the first term on the r.h.s is $\cos(ak) (ak)^{1 - \alpha} / (\alpha - 1)$. Secondly, since $x^{1- \alpha} \sin x \sim x^{2-\alpha}$ for $x \to 0$, the integral on the r.h.s converges to a finite value $C$ when $1 < \alpha < 3$ (note that it diverges for any $\alpha < 1$ and it is not defined for $\alpha=1$). Therefore we can establish that, for $1 < \alpha < 3$ : 
		\be 
			a^{\alpha - 1} \int_a^\infty dr ~ \frac{2 \cos(kr)}{r^\alpha} = \frac{2 \cos(ka)}{\alpha -1} + 2C (ak)^{\alpha - 1}
		\ee			
		for some constant $C$ (which depends on $\alpha$). Since $2 / (\alpha -1)$ is the value of the integral on the l.h.s at $k=0$, and since $\alpha - 1 < 2$, the dominant scaling at small $k$ in $d=1$ is : 
		\be 	
			\gamma_0 - \gamma_{\bm k} \sim k^{\alpha - 1} + O(k^2) ~~~~ (1 < \alpha < 3) ~.
		\ee
		If $\alpha > 3$, the integral on the r.h.s of Eq. \eqref{e.fourier_pot_1d} diverges at $k \to 0$ as $k^{3-\alpha}$, so that one obtains : 
		\be 
			a^{\alpha - 1} \int_a^\infty dr ~ \frac{2 \cos(kr)}{r^\alpha} = \frac{2}{\alpha -1} + C' (ak)^{2} + O(k^4)
		\ee			
		for some constant $C'$. Hence we can prove that in $d=1$ :
	\be 
		\gamma_0 - \gamma_{\bm k} \sim k^2 ~~~~ (\alpha > 3)  ~.
	\ee

	\subsection{$d=2$}		
	In $d=2$, the integral to be calculated is : 
	\begin{eqnarray}
		 a^{\alpha - 2} \int_a^\infty dr ~\int_0^{2\pi} d\theta ~ r^{1-\alpha} e^{ikr\cos \theta} \nonumber \\	
		  = 2 \pi (ak)^{\alpha - 2} \int_{ak}^\infty dx ~ x^{1-\alpha} {\cal J}_0(x)
	\label{e.fourier_pot_2d}
	\end{eqnarray}
	where ${\cal J}_n(x) = (1/2\pi) \int_0^{2\pi} e^{i(x\cos \theta - n\theta)} d\theta$ is a Bessel function. As ${\cal J}_0(0) = 1$, the integral on the r.h.s is divergent for $\alpha > 2$, and we perform an integration by parts. Using the fact that ${\cal J}_0'(x) = -{\cal J}_1(x)$, one has : 
	\begin{eqnarray}
		\int_{ak}^\infty dx ~ x^{1-\alpha} {\cal J}_0(x) & = & 
			\left[\frac{x^{2-\alpha}}{2-\alpha} {\cal J}_0(x) \right]_{ak}^\infty	\nonumber \\
			&+ & \int_{ak}^\infty dx ~ \frac{x^{2-\alpha}}{2-\alpha} {\cal J}_1(x) ~.
	\label{e.int_J1}
	\end{eqnarray}
	The first term on the r.h.s gives $(ak)^{2-\alpha}/(2-\alpha) + O(k^{4-\alpha})$, while, as ${\cal J}_1(x) \sim x$ at small $x$, the integral on the r.h.s converges when $2 < \alpha < 4$ and diverges for $\alpha > 4$ (note that as ${\cal J}_1(x) \sim \sqrt{2/\pi x} \cos(x-3\pi/4)$ for $x \to \infty$, the integral converges in $+\infty$). Thus, we have proved that, for $2 < \alpha < 4$ : 
	\be 
		2 \pi (ak)^{\alpha - 2} \int_{ak}^\infty dx ~ x^{1-\alpha} {\cal J}_0(x) = 
		  \frac{2\pi}{2-\alpha} + C k^{\alpha - 2} + O(k^2)
	\ee
	for some constant $C$. As $2 \pi / (2-\alpha)$ is the value of the integral on the l.h.s of Eq. \eqref{e.fourier_pot_2d} at $k=0$, we have thus established that in $d=2$ : 
	\be 
		\gamma_0 - \gamma_{\bm k} \sim k^{\alpha - 2} + O(k^2) ~~~ (2 < \alpha < 4) ~.
	\ee
	If $\alpha > 4$, the integral on the r.h.s of Eq. \eqref{e.int_J1} diverges as $k^{4-\alpha}$ when $k \to 0$, so that 
	\be 
		2 \pi (ak)^{\alpha - 2} \int_{ak}^\infty dx ~ x^{1-\alpha} {\cal J}_0(x) = 
		  \frac{2\pi}{2-\alpha} + C' k^{2} + O(k^3) 
	\ee
	for some constant $C'$. We have thus established that, in $d=2$ : 
	\be 
		\gamma_0 - \gamma_{\bm k} \sim k^{2} ~~~ (\alpha>4) ~.
	\ee
	
	\subsection{$d=3$}	
		In $d=3$, we have to evaluate : 
		\begin{eqnarray}
			a^{\alpha-3}\int_a^\infty dr ~\int_0^\pi d\theta ~2\pi r^2 \sin \theta \frac{e^{ikr\cos \theta}}{r^\alpha} \nonumber \\
			= 
			4\pi (ak)^{\alpha - 3} \int_{ak}^\infty dx ~ \frac{\sin x}{x^{\alpha-1}}  ~.
		\label{e.fourier_pot-3d}
		\end{eqnarray}	
		The integral on the r.h.s diverges in $k=0$ for $\alpha > 3$, and we integrate it by parts. The remaining integral shows the same divergence in $k=0$, so that a second integration by parts is needed. One obtains : 
		\begin{eqnarray}
			\int_{ak}^\infty dx ~ \frac{\sin x}{x^{\alpha-1}} & = &
				\frac{(ak)^{3-\alpha}}{\alpha-3} \\ 
				& - & \int_{ak}^\infty dx ~ \frac{x^{3-\alpha} \sin x}{(2-\alpha)(3-\alpha)} +
					 O(k^{5-\alpha}) ~.  \nonumber
		\label{e.fourier_pot-3d-2}
		\end{eqnarray}
		The first term on the r.h.s gives the $k=0$ value of the integral to evaluate in Eq. \eqref{e.fourier_pot-3d}. If $3 < \alpha < 5$, the integral on the r.h.s or Eq. \eqref{e.fourier_pot-3d-2}  converges, so that one obtains the scaling behavior in $d=3$ : 
		\be
			\gamma_0 - \gamma_{\bm k} \sim k^{\alpha - 3} + O(k^2) ~~~ (3 < \alpha < 5) ~.
		\ee
		If $5 < \alpha$, the integral on the r.h.s of Eq. \eqref{e.fourier_pot-3d-2} diverges in $k \to 0$ as $k^{5-\alpha}$, thus the result : 
		\be
			\gamma_0 - \gamma_{\bm k} \sim k^{2} ~~~ (\alpha > 5) ~.
		\ee

\section{Structure factor of the spin correlations}
\label{app-correl}
	In this Appendix, we detail all the calculations of the spin correlations in the ground state at the level of the spin-wave approximation. The starting point is the Holstein-Primakoff (HP) representation of the spin operators (we consider a general spin $s$) : 
		\bearr
			S_i^x & = &  s - b_i^\dagger b_i \\
			S_i^y & = & \frac{\sqrt{2s}}{2i}(b_i - b_i^\dagger) + O(b_i^3) \\
			S_i^z & = & -\frac{\sqrt{2s}}{2}(b_i + b_i^\dagger) + O(b_i^3)
	\eearr
	in the XY ferromagnetic phase (where the reference mean-field state has all spins pointing towards $+X$). In the N\'eel phase (where the reference state has spins pointing alternatingly towards $\pm Z$), we have : 
		\bearr
		S_i^z & = &  \epsilon_i (s - b_i^\dagger b_i ) \\
		S_i^y & = & \epsilon_i ~\frac{\sqrt{2s}}{2i}(b_i - b_i^\dagger) + O(b_i^3) \\
		S_i^x & = & \frac{\sqrt{2s}}{2}(b_i + b_i^\dagger) + O(b_i^3)
	\eearr
	where $\epsilon_i = +1$ on even sites, and $\epsilon_i = -1$ on odd sites. As the LSW approximate ground state, being the ground state of a quadratic Hamiltonian, satisfies Wick's theorem \cite{blaizot-ripka}, all the properties of the quantum fluctuations in the ground state are contained in the two-point correlation functions $\langle b_i^\dagger b_j\rangle$ and $\langle b_i b_j \rangle$. Higher order correlations can be expressed, through Wick's theorem, as a function of two-point ones \cite{blaizot-ripka}. If one assumes translational invariance, the correlations are most naturally expressed in momentum space. As the Bogoliubov rotation reads $b_{\bm k} = u_{\bm k} \beta_{\bm k} - v_{\bm k} \beta_{-\bm k}^\dagger$, and given the form of the Hamiltonian Eq. \eqref{e.H_quadra} in the main text, one obtains : 
	\bearr
		\langle b_{\bm k}^\dagger b_{\bm k'} \rangle &=& \delta_{\bm k, \bm k'} v_{\bm k}^2 \nonumber \\
		\langle b_{\bm k} b_{\bm k'} \rangle &=& - \delta_{\bm k, -\bm k'} u_{\bm k} v_{\bm k}
	\eearr
	where $v_{\bm k}^2 = (1/2)(A_{\bm k} / E_{\bm k} - 1)$ and $u_{\bm k} v_{\bm k} = B_{\bm k} / 2E_{\bm k}$. Here, $E_{\bm k} = \sqrt{A_{\bm k}^2 - B_{\bm k}^2}$, and $A_{\bm k}$ and $B_{\bm k}$ take the expressions in Eqs.~\eqref{e.AkBkXY} and \eqref{e.AkBkNeel}. 
	

Notice that none of the correlation functions possesses an imaginary part. An equivalent expression for the correlations in Fourier space is then:
	\bearr
		\langle b_{\bm k}^\dagger b_{\bm k} \rangle &=& \frac{1}{2}\left(\frac{A_{\bm k}}{E_{\bm k}} - 1 \right) \nonumber \\
		\langle b_{\bm k} b_{-\bm k} \rangle &=& - \frac{B_{\bm k}}{2 E_{\bm k}}
	\eearr	
The other correlations vanish.

\textit{XY phase.---} 
	In the XY phase, the rotational invariance about the $z$ axis is spontaneously broken. As a consequence, the $S^y S^y$ and $S^x S^x$ spin correlations need not have the same structure. At the level of the LSW approach, the $S^x S^x$ spin correlation is indeed of order 4 in the HP bosons operators, while the $S^y S^y$ correlation is only of order 2. One may thus expect that the decay exponent of the  $S^x S^x$ spin correlations is twice the decay exponent of the  $S^y S^y$ spin correlations, and this expectation is indeed confirmed numerically. The calculation for the  $S^x S^x$ spin correlations is better formulated in real space : 
	\bearr 	
		&& \langle S_i^x S_j^x \rangle - \langle S_i^x \rangle \langle S_j^x \rangle   \\
		&=& \langle (s - b_i^\dagger b_i)(s-b_j^\dagger b_j)\rangle - \langle s - b_i^\dagger b_i \rangle \langle s-b_j^\dagger b_j \rangle \nonumber \\
		&=& \langle b_i^\dagger b_i b_j^\dagger b_j \rangle -  \langle b_i^\dagger b_i \rangle \langle b_j^\dagger b_j \rangle 
		= \langle b_i^\dagger b_j^\dagger \rangle \langle b_i b_j\rangle + \langle b_i^\dagger b_j \rangle \langle b_i b_j^\dagger \rangle  ~. \nonumber
	\eearr
	where on the last line, we have used Wick's theorem.

	On the other hand, the  $S^y S^y$ spin correlations read (notice that $\langle S_i^y \rangle = 0$) : 
	\bearr 
		\langle S_i^y S_j^y \rangle &=& -\frac{s}{2} \langle (b_i - b_i^\dagger) (b_j - b_j^\dagger) \rangle  \\
		&=& -\frac{s}{2} \sum_{\bm k} [2 \langle b_{\bm k} b_{-\bm k} \rangle - 2 \langle b_{\bm k} b_{\bm k}^\dagger \rangle - 1] e^{i{\bm k}\cdot({\bm r}_i-{\bm r}_j)} \nonumber
	\eearr
	where we have used the fact that the correlation functions are real. One can directly deduce from this formula the expression of the static spin structure factor for the $y$ spin components: 
	\bearr
		S_{\bm k}^{yy} &=& \frac{s}{2} \sqrt{\frac{A_{\bm k} + B_{\bm k}}{A_{\bm k} - B_{\bm k}}} 
		= \frac{s}{2} \sqrt{\frac{1 - \Delta \gamma_{\bm k} / \gamma_0}{1 - \gamma_{\bm k} / \gamma_0}}
	\eearr
	Finally, the $S^z S^z$ correlations are: 
	\bearr 
		\langle S_i^z S_j^z \rangle &=& \frac{s}{2}  \langle (b_i +b_i^\dagger) (b_j + b_j^\dagger) \rangle  \\
		&=& \frac{s}{2} \sum_{\bm k} [2 \langle b_{\bm k} b_{-\bm k} \rangle + 2 \langle b_{\bm k} b_{\bm k}^\dagger \rangle + 1] e^{i{\bm k}\cdot({\bm r}_i-{\bm r}_j)}  \nonumber
	\eearr
	and we obtain the structure factor for the $z$ spin components as: 
	\bearr
		S_{\bm k}^{zz} &=& \frac{s}{2} \sqrt{\frac{A_{\bm k} - B_{\bm k}}{A_{\bm k} + B_{\bm k}}} 
		= \frac{s}{2} \sqrt{\frac{1 - \gamma_{\bm k} / \gamma_0}{1 - \Delta \gamma_{\bm k} / \gamma_0}}
	\eearr
	As $\sqrt{1 - \gamma_{\bm k} / \gamma_0} \sim k^z$ at small $k$, we thus obtain the scaling behavior for the structure factors in the XY phase : 
	\bearr
		S_{\bm k}^{yy} &\sim & k^{-z} \\
		S_{\bm k}^{zz} &\sim & k^z
	\eearr
	In correspondence with the small-$k$ behavior of the structure factors, we find that the associated correlations decay at large distance as : 
	\bearr
		\langle S_i^y S_j^y \rangle & \sim & 1/r^{d-z} \\
		\langle S_i^z S_j^z \rangle & \sim & 1/r^{d+z} ~.
	\eearr
	Indeed, if one assumes that correlations decay algebraically with distance, this behavior follows from the calculations of Appendix \ref{app-gamma_k}, where we proved that 
	\be 
		{\rm FT}\left[\frac{1}{x^{d+\sigma}} \right](k) \sim k^\sigma
	\ee
	as long as $-d < \sigma < 2$, and FT stands for Fourier transform.
	
	On the other hand, we verified numerically that the $S^x S^x$ spin correlations decay algebraically with a decay exponent twice as large as the decay exponent of the $S^yS^y$ correlations, as expected in view of the fact that the $S^xS^x$ correlation is quartic in the HP bosons, while $S^yS^y$ is quadratic: 
	\be 
		\langle S_i^x S_j^x \rangle - \langle S_i^x\rangle \langle S_j^x \rangle \sim 1/r^{2(d-z)}  ~.
	\ee

\textit{N\'eel phase.---} 
	In the N\'eel phase, the rotational symmetry about the $z$ axis is not broken, so that one expects $S_{\bm k}^{xx} = S_{\bm k}^{yy}$. Given the formula for the HP transformation, the expression of $S_{\bm k}^{xx}$ in the N\'eel phase is exactly the same as the expression of $S_{\bm k}^{zz}$ in the XY phase, namely : 
	\be 
		S_{\bm k}^{xx} = \frac{s}{2} \sqrt{\frac{A_{\bm k} - B_{\bm k}}{A_{\bm k} + B_{\bm k}}}
	\ee
	On the other hand, the calculation for the $S^y S^y$ spin correlations gives (with ${\bm K} = (\pi, \pi, \dots)$) : 
	\bearr 
		\langle S_i^y S_j^y \rangle &=& -\frac{s}{2} e^{i{\bm K}\cdot({\bm r}_i - {\bm r}_j)} \langle (b_i - b_i^\dagger) (b_j - b_j^\dagger) \rangle  \\
		&=& -\frac{s}{2} \sum_{\bm k} [2 \langle b_{\bm k} b_{-\bm k} \rangle - 2 \langle b_{\bm k} b_{\bm k}^\dagger \rangle - 1] e^{i({\bm k-\bm K})\cdot(\bm r_i- \bm r_j)}  \nonumber
	\eearr
	from which we can deduce the structure factor :
	\be 
		S_{\bm k}^{yy} = \frac{s}{2} \sqrt{\frac{A_{{\bm k}-{\bm K}} + B_{{\bm k}-{\bm K}}}{A_{{\bm k}-{\bm K}} - B_{{\bm k}-{\bm K}}}}
	\ee
	
	Given the expression of $A_{\bm k}$ and $B_{\bm k}$ in the N\'eel phase, we thus have $S_{\bm k}^{xx} = S_{\bm k}^{yy}$, as expected. The structure factor scales as the dispersion relation at small momentum: 
	
	\be 
		S_{\bm k}^{xx} = S_{\bm k}^{yy} \sim \Delta + k^{2z}
	\ee
	
	 From the expression of the structure factor, one can then reconstruct the spatial decay of the spin correlations. We find an exponential decay at short distance (associated with the gap $\Delta$), and an algebraic decay at large distance, with an exponent $\alpha$, directly controlled by the decay of the spin-spin interaction. Finally, the $S^z S^z$ spin correlations read: 
	 
	\be
		\langle S_i^z S_j^z \rangle - \langle S_i^z \rangle \langle S_j^z \rangle = e^{i\bm K \cdot (\bm r_i -\bm r_j)}[\langle b_i^\dagger b_j^\dagger \rangle \langle b_i b_j\rangle + \langle b_i^\dagger b_j \rangle \langle b_i b_j^\dagger \rangle]
	\ee	 	
	which also decays as $1/| \bm r_i - \bm r_j |^\alpha$ at large separation, and show a staggered pattern.  }

\section{Scaling of the covariance}
\label{app-scaling-covar}
\begin{figure}
		\includegraphics[width=0.6\linewidth]{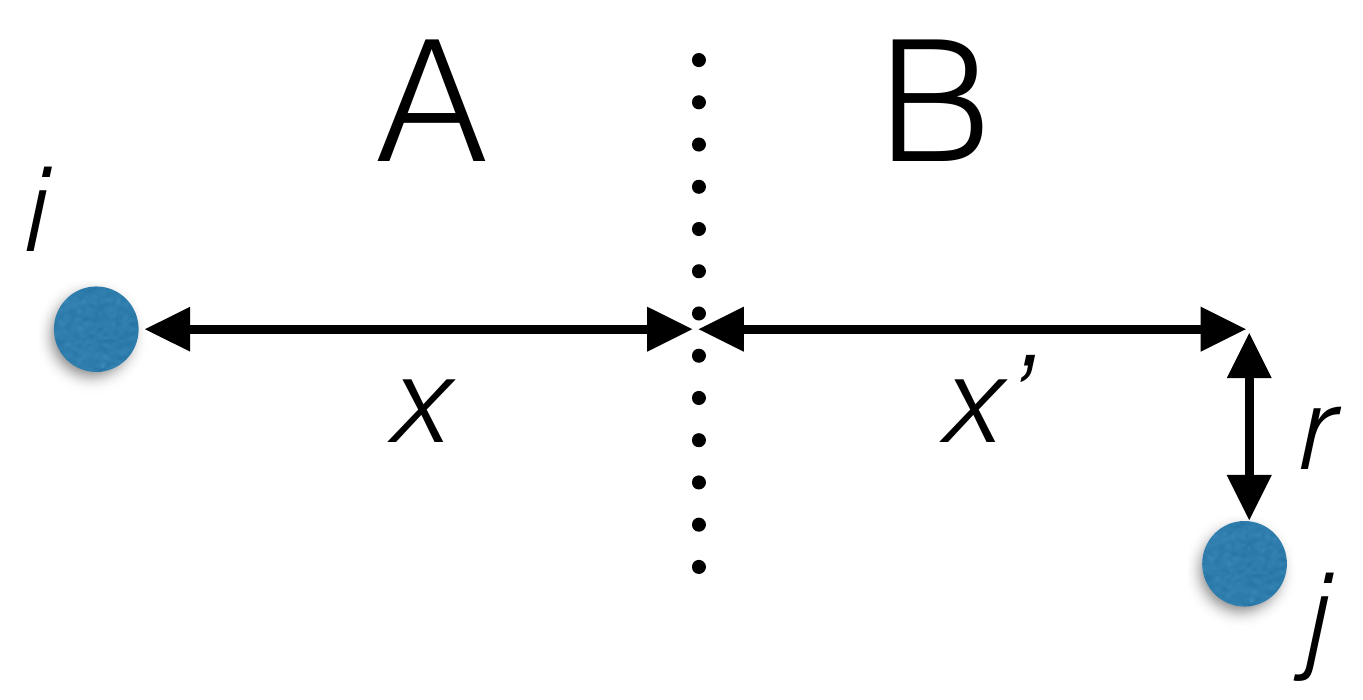}
		\caption{Geometry of the variables used in the integration leading to the scaling of the covariance  $\langle \delta {\cal O}_A \delta{\cal O}_B \rangle$.}
		\label{f.schema_contour}
\end{figure}

In this Appendix, we provide details on the calculation leading to the prediction of Eq. \eqref{eq_scaling_covar}. Considering two subsystems $A$ and $B$ of linear size $L$ and two observables ${\cal O}_{A/B} = \sum_{i \in A/B} {\cal O}_i$, the covariance of ${\cal O}_A$ and ${\cal O}_B$ is given by
\bearr
	\langle \delta {\cal O}_A \delta{\cal O}_B \rangle &=& \langle {\cal O}_A {\cal O}_B \rangle -  \langle {\cal O}_A \rangle \langle {\cal O}_B \rangle \\
	&=& \sum_{i \in A} \sum_{j\in B} \langle \delta {\cal O}_i \delta{\cal O}_j \rangle~.
\eearr 
Assuming that the correlation function decays as a power-law with distance : 
\be 
	\langle \delta {\cal O}_i \delta{\cal O}_j \rangle \sim \frac{1}{\vert {\bm r}_i - {\bm r}_j \vert^\eta}
\ee
we can relate the scaling behavior of $\langle \delta {\cal O}_A \delta{\cal O}_B \rangle$ to the exponent $\eta$. For the sake of mathematical simplicity, we take $A$ and $B$ translationally invariant in $d-1$ dimensions (they are thus lines in $d=1$ and cylinders in $d=2$, while in $d=3$, the geometry is more difficult to visualize, since the $A-B$ contact area has the topology of a torus. As our aim is simply to predict the scaling of the covariance with the linear size of $A$ and $B$, their precise shape is irrelevant. The analysis proceeds in two steps 1) we analyse the decay of the \textit{correlation contour} \cite{FrerotR2015} 
\be 
	C_i = \sum_{j \in B} \langle \delta {\cal O}_i \delta{\cal O}_j \rangle
\ee
where $i$ is a site in $A$, whose distance from the $A-B$ boundary is denoted $a$, as illustrated on Fig. \ref{f.schema_contour};

and 2) we deduce the scaling of $\langle \delta {\cal O}_A \delta{\cal O}_B \rangle = \sum_{i \in A} C_i$ from the decay of $C_i$ when moving away from the $A-B$ boundary. 
\\
\textit{Decay of the contour.---}
We have to evaluate the scaling behavior of 
\bearr
	C(x) &\sim & \int_0^L dx' \int_0^L dr~ \frac{r^{d-2}}{[r^2 + (x+x')^2]^{\eta/2}} \nonumber \\
	&\sim & x^{d-\eta} \int_0^{L/x} du~ (1+u)^{d-1-\eta} \nonumber \\
	&& \times \int_0^{(L/x)/(1+u)} \frac{dv~ v^{d-2}}{(1+v^2)^{\eta/2}}
\eearr
The last integral converges if $\eta \ge d-1$ and is of order $[(L/x)/(1+u)]^{d-1-\eta}$ otherwise (with logarithmic corrections if $\eta = d-1$). In this latter case ($\eta < d-1$), we find that $C(x) \sim L^{d-\eta}$. If $\eta > d-1$, we have that
\be 
	C(x) \sim x^{d-\eta} \int_0^{L/x} du ~(1+u)^{d-1-\eta}
\ee
If $\eta > d$, the integral converges and $C(x) \sim x^{d-\eta}$. If $\eta < d$, the integral is of order $(L/x)^{d-\eta}$ and we find again that $C(x) \sim L^{d- \eta}$. We also predict logarithmic corrections if $\eta = d$. To sum up, we have the following behavior for the contour
		\be
			C(x) = \sum_{j \in B} \frac{1}{| {\bm r}_i - {\bm r}_j |^{\eta}} \sim 
			 \left\{
      \begin{aligned}
        L^{d - \eta} & ~~\textnormal{if} & \eta < d \\
        x^{d-\eta} & ~~\textnormal{if} & \eta > d \\
      \end{aligned}
    \right.
    \label{eq_scaling_contour}
		\ee	
with logarithmic corrections if $\eta =d$ or $\eta=d-1$. 
\\
\textit{Scaling of the covariance.---}
Summing the contour $C_i$ over all sites in $A$ gives a trivial factor of $L^{d-1}$ for the $d-1$ directions transverse to $x$. Integrating the distance $x$ to the $A-B$ boundary gives the final result of Eq. \eqref{eq_scaling_covar} with further logarithmic corrections if $\eta = d$ or $\eta = d-1$.
\bibliography{biblio_LR,biblio_EE}

\end{document}